\documentclass[%
preprint,
 amsmath,amssymb,
 aps,
]{revtex4-2}

\usepackage{epstopdf} 

\usepackage{graphicx}

\usepackage{dcolumn}
\usepackage{bm}
\usepackage{hyperref}

\usepackage{epstopdf}
\usepackage{lineno}
\usepackage{textcomp}
\usepackage[gen]{eurosym}
\usepackage{slashed}
\usepackage[dvipsnames]{xcolor}  
\usepackage{placeins}

\usepackage{fancyheadings}
\usepackage{lastpage}
\usepackage{caption} 
\usepackage{verbatim}
\usepackage{color} 
\usepackage{subcaption} 
\usepackage{datetime}
\usepackage{booktabs, tabularx, multirow, array} 
\usepackage[separate-uncertainty,retain-explicit-plus,per-mode=symbol,binary-units]{siunitx}
\usepackage[version=4]{mhchem}
\DeclareSIUnit\c{\mbox{$c$}}
\DeclareSIUnit\year{yr}

\begin{document}


\title{ \textbf{ {\LARGE A  low-mass dark matter project, \textcolor{DarkOrchid}{ALETHEIA}: \textcolor{DarkOrchid}{A} \textcolor{DarkOrchid}{L}iquid h\textcolor{DarkOrchid}{E}lium \textcolor{DarkOrchid}{T}ime projection c\textcolor{DarkOrchid}{H}amb\textcolor{DarkOrchid}{E}r \textcolor{DarkOrchid}{I}n d\textcolor{DarkOrchid}{A}rk matter} } }

\thanks{Corresponding authors: Junhui Liao and Yuanning Gao.}%

\author{Junhui Liao}
\email{junhui\_liao@brown.edu, junhui\_163@163.com}
\affiliation{%
Division of Nuclear Physics, China Institute of Atomic Energy, Beijing, China 
}%
\affiliation{
Department of Physics , Brown University
}

\author{Yuanning Gao}
\email{yuanning.gao@pku.edu.cn}
\affiliation{%
 School of Physics, Peking University 
}%

\author{Zhuo Liang, Chaohua Peng, Jian Zheng, Jiangfeng Zhou}
\affiliation{%
The Division of Nuclear Physics , China Institute of Atomic Energy, Beijing, China 
}%

\author{Fengshou Zhang, Lei Zhang}
\affiliation{%
The Division of Nuclear Synthesis Technology, China Institute of Atomic Energy, Beijing, China 
}%

\author{Zebang Ouyang}
\affiliation{%
The School of Nuclear Technology, University of South China, Hengyang, Hunan, China 
}%


\date{\today}

\begin{abstract}
Dark Matter (DM) is one of the most critical questions to be understood and answered in fundamental physics today. Observations with varied astronomical and cosmological technologies already pinned down that DM exists in the Universe, the Milky Way, and the Solar System. However, the understanding of DM under the language of elementary physics is still in progress. DM direct detection aims to test the interactive cross-section between galactic DM particles and an underground detector's nucleons. Although Weakly Interactive Massive Particles (WIMPs) is the most discussed DM candidate, the null-WIMPs conclusion has been consistently addressed by most convincing experiments in the field. The low-mass WIMPs region (100s MeV/c$^2$ - 10 GeV/c$^2$) has not been fully exploited comparing to high-mass WIMPs (10 GeV/c$^2$ - 1 TeV/c$^2$) experiments which implement liquid xenon or argon TPCs (Time Projection Chambers). The ALETHEIA experiment aims to hunt for low-mass WIMPs with liquid helium-filled TPCs. In this paper, we go through the physics motivation of low-mass DM, the ALETHEIA detector's design, a series of R\&D programs that should be launched to address a liquid helium TPC's functionality, and possible analysis channels available for DM searches.
\end{abstract}

\maketitle

\tableofcontents


\section*{The existence of dark matter and its detection} 

Astronomical evidence of many types, including cluster and galaxy rotation curves~\cite{Zwicky33, Rubin70}, lensing studies and spectacular observations of galaxy cluster collisions~\cite{Refreiger03,Clowe06,Fields08}, and cosmic microwave background (CMB) measurements~\cite{Planck2018ResultsOne}, all point to the existence of cold dark matter (CDM) particles.  Cosmological simulations based on the CDM model have been remarkably successful at predicting the actual structures we see in the Universe.  Alternative explanations involving modification of general relativity have not been able to explain this large body of evidence across all length scales~\cite{feng2014planning}. 

Recent results from Gaia~\cite{GaiaExp} showing consistency with many previous experiments therefore much more securely pinned down than ever on: (a) DM dominates the mass of the Milky Way Galaxy~\cite{PostiHelmi19}, and (b) the local DM mass density in the Solar system is around 0.3 GeV/c$^2$ $\cdot$ cm$^{-3}$~\cite{HagenHelmi18}.

Weakly Interactive Massive Particles (WIMPs) and Axions are among the two most prominent  DM candidates.  WIMPs are a hypothesized class of DM particles that would freeze out of thermal equilibrium in the early Universe with a relic density that matches observation~\cite{Feng10}.  The so-called ``WIMP miracle'' is the coincident emergence of WIMPs with similar characteristics both from the solution of the gauge hierarchy problem and from the observed relic density of dark matter.
Axions are motived by the strong Charge Parity (CP) problem, essentially to solve a fine-tuning problem of 1 part in $\sim$ 10$^{10}$~\cite{Feng10}.  Axions are much lighter and require different experimental techniques than WIMPs to detect them as in the Axion Dark Matter eXperiment (ADMX)~\cite{ADMX14}. The KSVZ and DFSZ models have been excluded for axions masses between [2.66, 3.31]~$\mu$eV with 90\% confidence level~\cite{ADMXPRL20}. 

There are several viable strategies to detect DM.  Indirect detection experiments aim to observe high-energy particles resulting from the self-annihilation of DM.  Collider experiments look for the production of DM particles in high energy collisions.  Direct detection experiments aim to observe the rare scatters of DM on the low threshold, very low background detectors operated in deep underground Laboratories. Table~\ref{tabDMExpCat} categorizes currently active DM experiments.

\newcommand{\otoprule}{\midrule[\heavyrulewidth]} 

\begin{table}[ht]
   \centering
   \caption{Categorization of DM experiments with physics motivation} \label{tabDMExpCat}   
      \begin{tabular}{c c c c}
      \toprule%
         \multicolumn{1}{c}{\bfseries{ low-mass } }  &
         \multicolumn{1}{c}{\bfseries{ high-mass} }  &
         \multicolumn{1}{c}{\bfseries{Annual modulation} }  &
         \multicolumn{1}{c}{\bfseries{Directional} }     \\%
	  \otoprule%
      ALETHEIA, CEDX,				&ArDM, DarkSide, 		&ANAIS,				& DRIFT, \\
      CRESST, DAMIC, Edelweiss,		&DEAP, LZ, PandaX, 	&COSINE-100, 		& NEWAGE, \\
      SENSI, SuperCDMS.				&PICO, XENON.	 	&DAMA/LIBRA.			& NEWS. \\
      \bottomrule
    \end{tabular}
\end{table}

This proposal will focus on the \textit{ALETHEIA} project for the experimental detection and verification of the WIMPs hypothesis.

\subsection*{The context of  low-mass WIMPs direct detection}

The lowest limit for WIMPs-nucleon interaction for $\sim$ 30 GeV/c$^2$ is down to $\sim$ 10$^{-47}$ cm$^2$~\cite{Xenon1TPaper18}, which is eight orders lower than the weak interaction through the Z boson ($\sim$ 10$^{-39}$ cm$^2$). Assuming the WIMPs-nucleon scattering is through the Higgs boson, the cross-section would be $\sim$ 10$^{-46}$ cm$^2$, which is the lowest ``natural'' cross-section for weak interaction and has also been excluded by Xenon-1T data on the WIMPs mass region of $\sim$ 10 GeV/c$^2$ - 50 GeV/c$^2$. Taking 35 GeV/c$^2$ WIMPs as an example, with the cross-section of 10$^{-46}$ cm$^2$ and the exposure of 1 ton * yr, the expected events in XENON-1T are $\sim$ 21~\footnote{The 21 events were under the assumption that the acceptance of XENON-1T among its ROI nuclear recoil energy, 4.9 - 40.9 kev$_{nr}$, is 100\%. If considering the acceptance, the number of events is estimated to be slightly less than 21.}, while the observed DM candidate events are $\sim$ 0~\cite{Xenon1TPaper18}. The Xenon-1T results ruled out that the mediator of WIMPs-nucleon interaction is the Higgs boson for WIMPs mass greater than $\sim$ 10 GeV/c$^2$. 
However, in theory, the WIMPs scenario could still be ``long-life'', as mentioned in reference~\cite{HooperBrownTalk2018}. Among other possibilities, one way to reconcile the conflict, WIMPs are supposed to be detected but failed, is dark matter might turn out to be lighter than the mass region where  high-mass experiments (DarkSide, LUX/LZ, PandaX, and Xenon) can effectively reach. For instance, $\sim$ 100s MeV/c$^2$ - 10 GeV/c$^2$. In this note below, unless specially mentioned, we will consider $\sim$ 100s MeV/c$^2$ - 10 GeV/c$^2$  as a default  for low-mass DM. 
In fact, independent of SuperSymmetry (SUSY) models which predict weakly mass WIMPs~\cite{Jungman96}, there are four categories of well-motivated scenarios favor MeV/c$^2$ - GeVc/c$^2$ dark matter:  (a), the ``WIMPs miracle'' model motivates $\sim$ 10 MeV/c$^2$ - 100 TeV/c$^2$ WIMPs~\cite{Feng10, FengKumar08}: a stable particle with such mass annihilating each other with a cross-section of $\sigma v \sim 2 \times 10^{-26} cm^3/s$ in the early university would result in consistent dark matter density as measured~\cite{Steigman12}. DM mass greater than 120 TeV/c$^2$ would violate the requirement of partial wave unitarity~\cite{Griest90}, while the Big Bang nucleosynthesis would ruin if an annihilating relic with a mass lighter than $\sim$ 1-10 MeV/c$^2$~\cite{Bohm13}.
(b), Light DM annihilates to quarks might have been suppressed during the epoch of recombination; instead of coupling to quarks, light DM would possibly couple to leptons, in particular, electrons. In direct detection, DM could scatter with electrons to generate individual electrons, individual photons, individual ions, and heat/phonons~\cite{Essig12}. 
(c), The asymmetry in the dark sector might be related to the baryon asymmetry (matter and anti-matter) ,  resulting in zero net baryon number of the universe ~\cite{Cohen93, Zurek14}. The mass of DM would be $\sim r \times (4-5)$ GeV/c$^2$, where $r$ is a factor that maintains equilibrium between the dark and SM (Stand Model) sectors in the early universe. If $r = 1$, the DM mass would be $\sim (4-5)$ GeV/c$^2$. 
(d), Strongly Interacting Massive particle (SIMP) models proposes dark matter as a meson- or baryon-like bound-state of hidden sector particles, with a mass near the QCD scale, $\sim$ 100 MeV/c$^2$~\cite{StrasslerZurek07, Hochberg14, Hochberg15,Kuflik16,Kuflik17}. Considering constraints from the CMB, dark matter masses might be the range of $\sim$ 5 - 200 MeV/c$^2$.

Regardless of whether or not WIMPs could be discovered at the scale of $\sim$ 100 GeV/c$^2$,  low-mass WIMPs should still be investigated, since (1) it has been motivated by several different mechanisms, as mentioned above; (2) considering there exist tens of element particles in the Standard Model, it's reasonable to guess the existence of more than one dark matter particle, which would naturally have different masses. So, even WIMPs were discovered in the  high-mass region, physicists should also check  low-mass region to see if there exist  low-mass dark matter there, and vice versa.

However, as mentioned, none of the leading high-mass DM experiments is sensitive to  low-mass WIMPs due to their relative heavy target material, xenon, or argon~\footnote{I am focusing on the traditional Spin-Independent model with "S1/S2" analysis. By applying other analysis methods like S2O and new models like Migdal-effect, these liquid heavy noble gas experiments could be sensitive to WIMPs mass down to the sub-GeV region. We will address this in detail below.}, though these experiments could achieve extremely low backgrounds. For instance, the 10-ton total mass LZ would have only $\sim$ six background events with 16.8 ton*yr exposure~\cite{LZTDR17}, the 50 kg DarkSide-50 detector demonstrated a background-free search with an exposure of 16.7 ton*day~\cite{DarkSide502018}.

The ALETHEIA was inspired by both high-mass and low-mass WIMPs experiments in the community. Although there existed quite a few low-mass DM experiments~\cite{SuperCDMS2014, SuperCDMS2016, DAMIC2016, CRESST2014, CRESST2017, EDELWEISS2017, CDEX2018}, we believe the ALETHEIA is a competitive project in the race of hunting for low-mass DM thanks to the following advantages. \\
(a) Being able to discriminate ER/NR events with "S1/S2" and possibly PSD ( Pulse Shape Discrimination  ) technologies. \\
(b), LHe could achieve extremely low or even zero intrinsic backgrounds. At 4 K temperature, $^3$He is the only solvable material  in LHe while it is very rare in Nature; any other impurities would show in the solid-state. Impurities are supposed to be purified completely with getter and cold-trap technologies. \\
(c), $^4$He only has two nucleons and four electrons. Therefore the intrinsic ER background induced by (background) gammas and neutrinos would be significantly smaller than heavier noble elements like Ar and Xe. \\
(d), TPC is a technology demonstrated for LXe and LAr; it could possibly be employed on LHe. If an LHe TPC were made successfully, its fiducial volume is expected to achieve very few or even zero backgrounds by self-shielding and other data-analyzing technologies. \\
and (e), being capable of scaling up to a ton or multi-tons size to fully touch down the B-8 neutrino floor. \\

Using superfluid helium, reference~\cite{Hertel19}, and~\cite{Maris17} search for $\sim$ MeV/c$^2$ DM with Transition Edge Sensor (TES) and ionization field, respectively.

\section*{Introduction to the ALETHEIA experiment}
\begin{figure}[!t]	 
	\centering
        \includegraphics[width=6.5in, angle = 0]{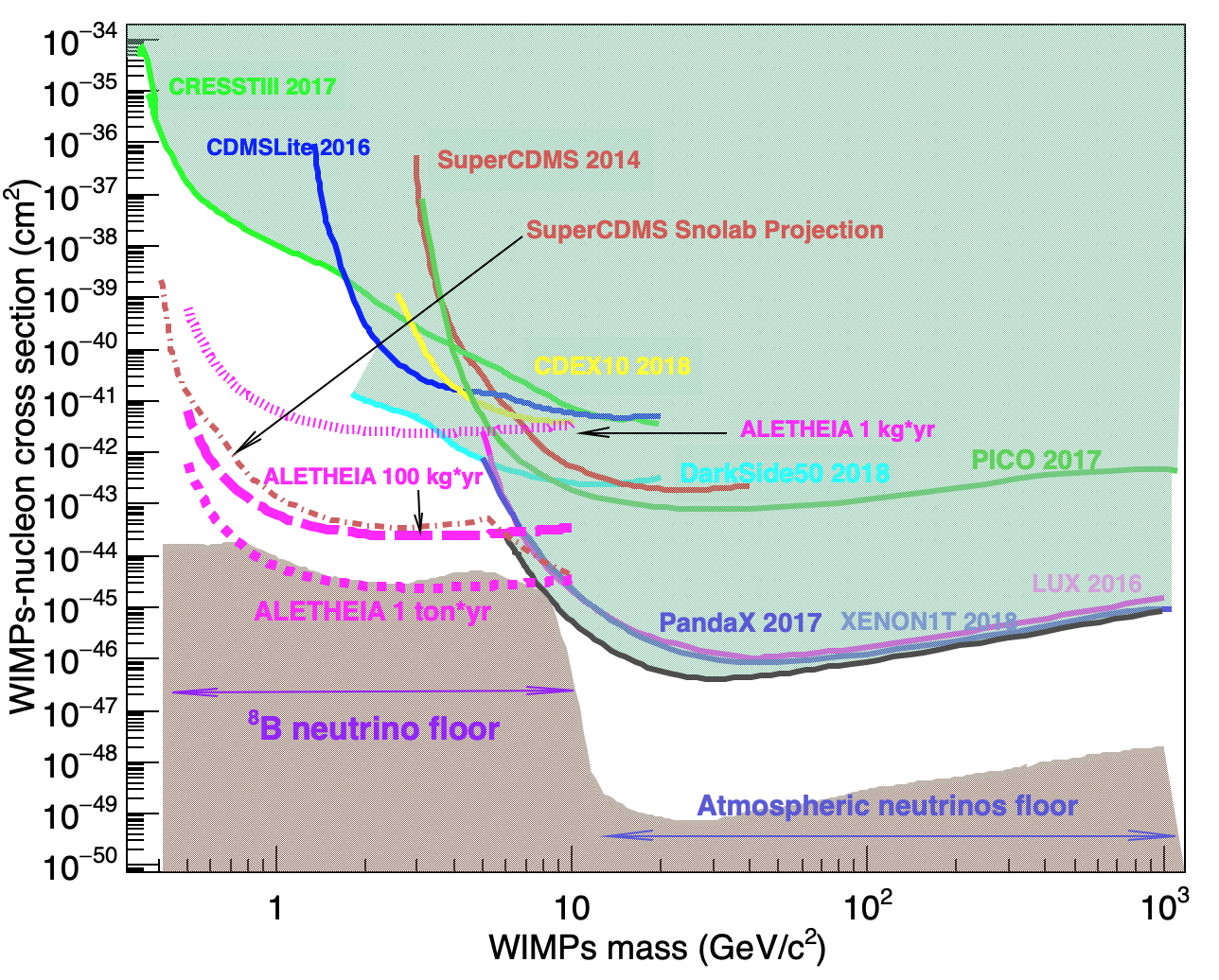}
	\caption{The figure shows the parameter space for the spin-independent WIMPs-nucleon cross-section. The dark green region represents the space that has been excluded by leading direct detection experiments at a 90\% confidence level. The area with light brown color corresponds to the space of the ``neutrino floor'' where neutrinos can generate WIMPs-like events in DM detectors. The projected sensitivities of the ALETHEIA with the exposure of 1 kg*yr, 100 kg*yr, and 1 ton*yr are shown. The upper limits or projected limits of other experiments are also show. The ``CRESSTIII-2017'' data cited from~\cite{CRESST2017}, ``CDMSLite-2016'' from~\cite{CDMSLite2016}, ``CDEX10 2018'' from~\cite{CDEX2018}, ``DarkSide50 2018'' from~\cite{DarkSide502018}, ``LUX2016'' from ~\cite{LUXCompleteExp2016}, ``PandaX 2017'' from~\cite{PandaX2017}, ``PICO 2017'' from~\cite{PICO2017}, ``Neutrino floor'' from~\cite{Billard14}., ``SuperCDMS 2014'' from~\cite{SuperCDMS2014}, ``SuperCDMS Snolab projection'' from~\cite{SuperCDMS2016},  ``XENON1T 2018'' from~\cite{Xenon1TPaper18}.}\label{projectedSensitivities} 
\FloatBarrier
\end{figure}

The ALETHEIA project is an instrumental background-free experiment with the ROI (Research Of Interest) of $\sim$ 100s MeV/c$^2$ - 10 GeV/c$^2$. Fig~\ref{projectedSensitivities} shows the projected sensitivity of the ALETHEIA with the exposure of 1 kg*yr, 100 kg*yr, and 1 ton*yr, respectively. The projected sensitivity is based on two assumptions: (a), there is no any background in the interesting recoil energy range (after a series of cuts being applied), 0.5 keV$_{nr}$ - 10 keV$_{nr}$; (b), the detection efficiency is 100\% for the whole ROI energy interval. The detection efficiency has not been tested by any group yet. We will test it with our apparatuses, which are currently under construction. 

With the S2 only analysis, the ALETHEIA project could be sensitive to 10s MeV/c$^2$ WIMPs. Because the analysis depends on data, so it is not possible to project the sensitivity for 10s MeV/c$^2$ WIMPs at this stage, for details, see below, please. 

Here, instrumental background-free means in the range of ROI, a small number of background events (for instance, $<$ 0.1 events) are expected, such that zero background events are observed. The instrumental backgrounds include radioactive particles due to the materials in the detector system (including dust) and the particles generated by cosmological muons hitting the rocks near the detector or the detector itself. For  low-mass WIMPs searches, there is another background which is registered by $^8$B solar neutrinos. The $^8$B events can't be discriminated from  low-mass WIMPs signals. According to reference~\cite{Bahcall04, Billard14, SNO02}, the measured $^8$B events are well consistent with the theoretical prediction. The uncertainty of the events is $\sim$16\%, as reported in reference~\cite{SNO02}. Consequently, in a WIMPs detector that is free of instrumental background like the ALETHEIA, by subtracting (theoretically estimated) $^8$B events from the observed events, we can figure out how many WIMPs events were observed and this quantity's uncertainty.

\section*{The legacy of LAr TPC experiments in DM direct detection}

1984 Noble physics laureate Prof. Carlo Rubia has firstly proposed the concept of Time Projection Chamber (TPC) in 1977~\cite{Rubia77} and implemented in the ICARUS experiment for the research of neutrinos filled with Liquid Argon (LAr)~\cite{ICARUS88}. In 1993, the ICARUS R\&D program shown the LAr TPC was capable of discriminating heavy-ion recoils from $\beta$ and $\gamma$ radiation~\cite{ICARUS93}. As a spin-off of ICARUS, WARP was the first experiment utilizing the LAr TPC with 2.6 kg active mass to hunt for dark matter~\cite{WARP08}. The Pulse Shape Discrimination (PSD) technology was implemented into the WIMPs search.

\begin{figure}	
	\centering
	\begin{subfigure}[t]{2.5in}
		\centering
		~~~\includegraphics[scale=0.35]{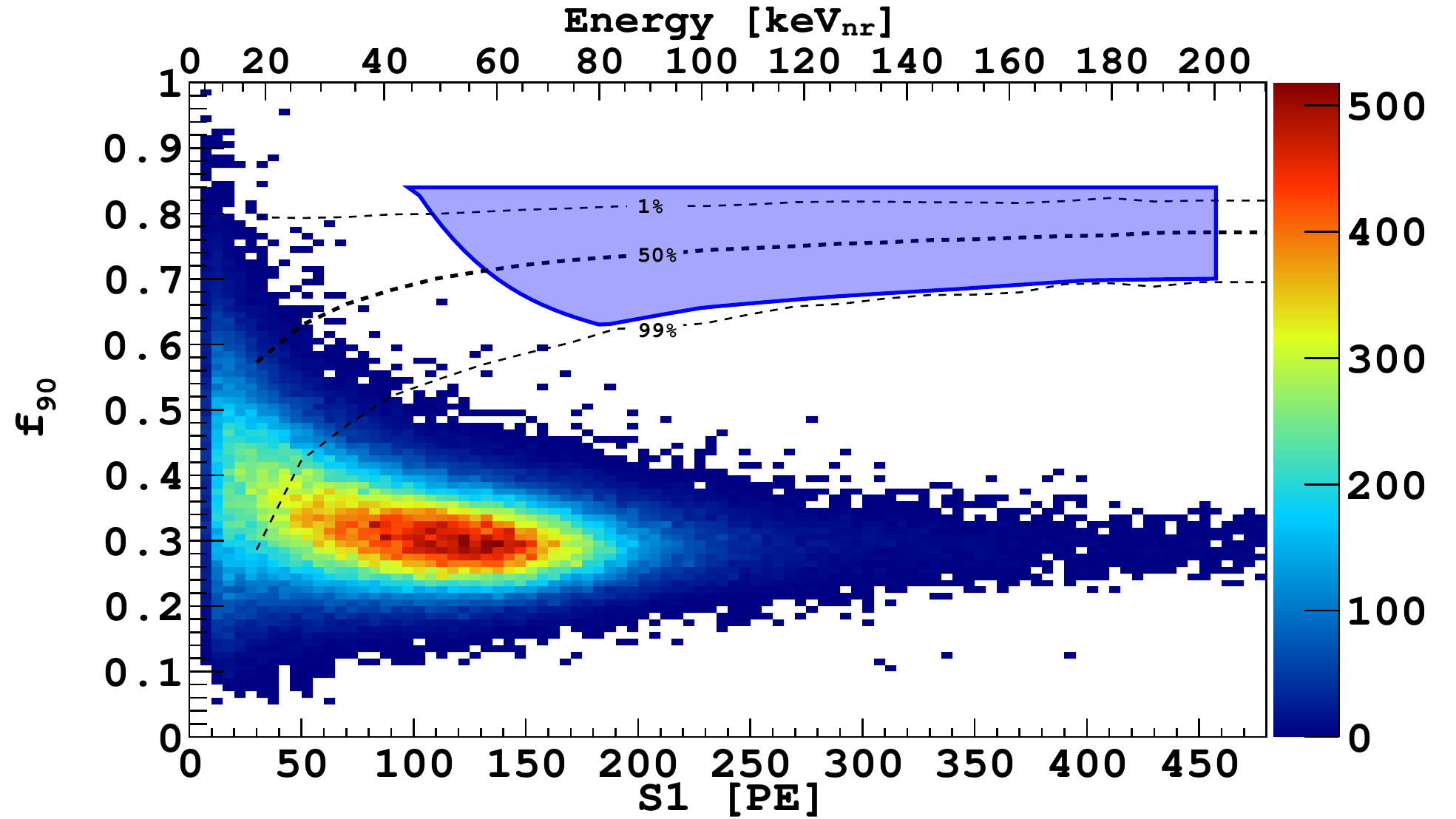}
		\caption{Fig.~\ref{fig_2.a}. The analysis uses PSD only.}\label{fig_2.a}	
	\end{subfigure}
	\begin{subfigure}[t]{3.8in}
		\centering
		\includegraphics[scale=0.35, angle = 0]{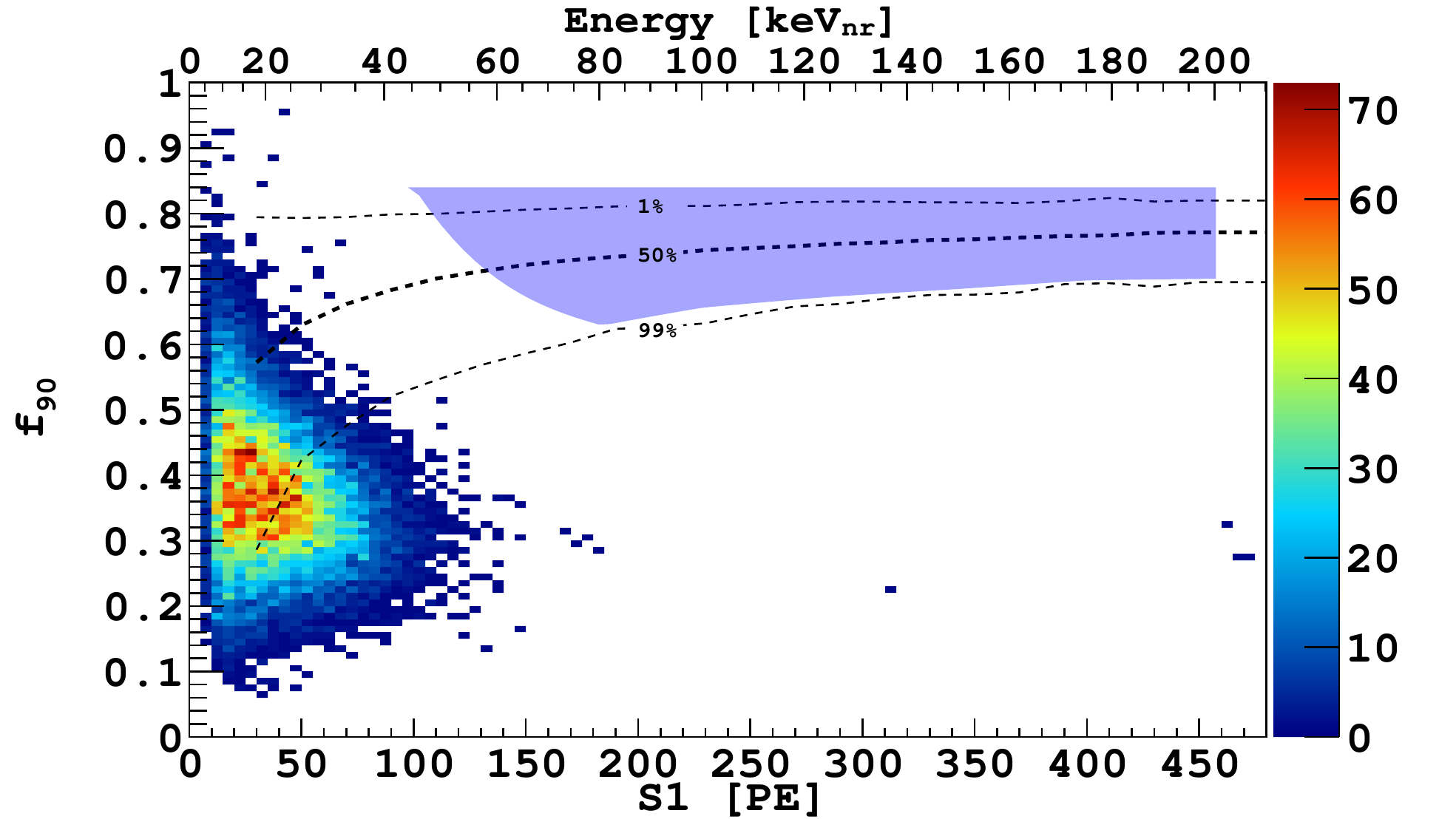}
		\caption{Fig.~\ref{fig_2.b}. The analysis uses both PSD and ``S1/S2''.}\label{fig_2.b}
	\end{subfigure}
	\caption{Survived events on the plane of f$_{90}$ vs. S1 after all of the cuts been applied with the exposure of 532 days of DarkSide-50~\cite{DarkSide502018}. On both sub-figures, the population of the event on the left lower corner represented ER backgrounds, and the contour in light blue corresponded to the ROI of DarkSide-50. It is easy to see that there exist zero events inside of the ROI in which less than 0.1 events were expected. This feature is a clear demonstration of instrumental background free. For the analysis shown in Fig.~\ref{fig_2.a}, only PSD was used. For the analysis on Fig.~\ref{fig_2.b}, both PSD and ``S1/S2'' were implemented. The two plots are copied from reference~\cite{DarkSide502018}.}\label{fig_2}
\FloatBarrier
\end{figure}

Using $\sim$ 46.4 kg LAr, DarkSide-50 experiment demonstrated a background-free state has been achieved in the energy range of $\sim$[40, 200] keV$_{\text{nr}}$~\cite{DarkSide502015, DarkSide502018}, as shown in Fig.~\ref{fig_2.a}.  
To achieve the feature of background-free, among other ``common'' techniques widely utilized in DM direct detection experiments with a TPC, like using outer detectors as a veto to reject neutron backgrounds and fiducial volume cuts to further remove wall and surface events etc, the critical technique used to achieve the background-free state of the DarkSide-50 experiment is PSD. For Electron Recoil (ER) and Nuclear Recoil (NR) induced scintillations in LAr, the ratio of scintillation in the first 90 ns and 7 $\mu$s, $f_{90}$ defined as \textit{(charge of signals in the first 90 ns)} / \textit{(charge of signals in the first 7 $\mu$s)}, is different. Typically, the $f_{90}$ for ER events is $\sim$ 0.3, while for NR events is $\sim$ 0.7. Reference~\cite{DarkSide502015} shows that PSD is capable of entirely discriminating all 1.5$\times$10$^7$ Ar-39-induced ER events from NR events (to achieve its background-free state). 

As addressed by DarkSide-50~\cite{DarkSide502018}, by combining the technique of PSD with ``S1/S2'' (which is the method used to discriminate ER/NR in LXe TPCs), the capability of ER/NR discrimination can be improved significantly. As shown in Fig.~\ref{fig_2.b}, comparing to Fig.~\ref{fig_2.a}, the background events were significantly constrained thanks to the implementation of ``S1/S2'', which could provide extra background constraint. 

\captionsetup[subfigure]{labelformat=empty}

\begin{figure}	
	\centering
	\begin{subfigure}[t]{3.0in}
		\centering
		~~~\includegraphics[scale=0.4]{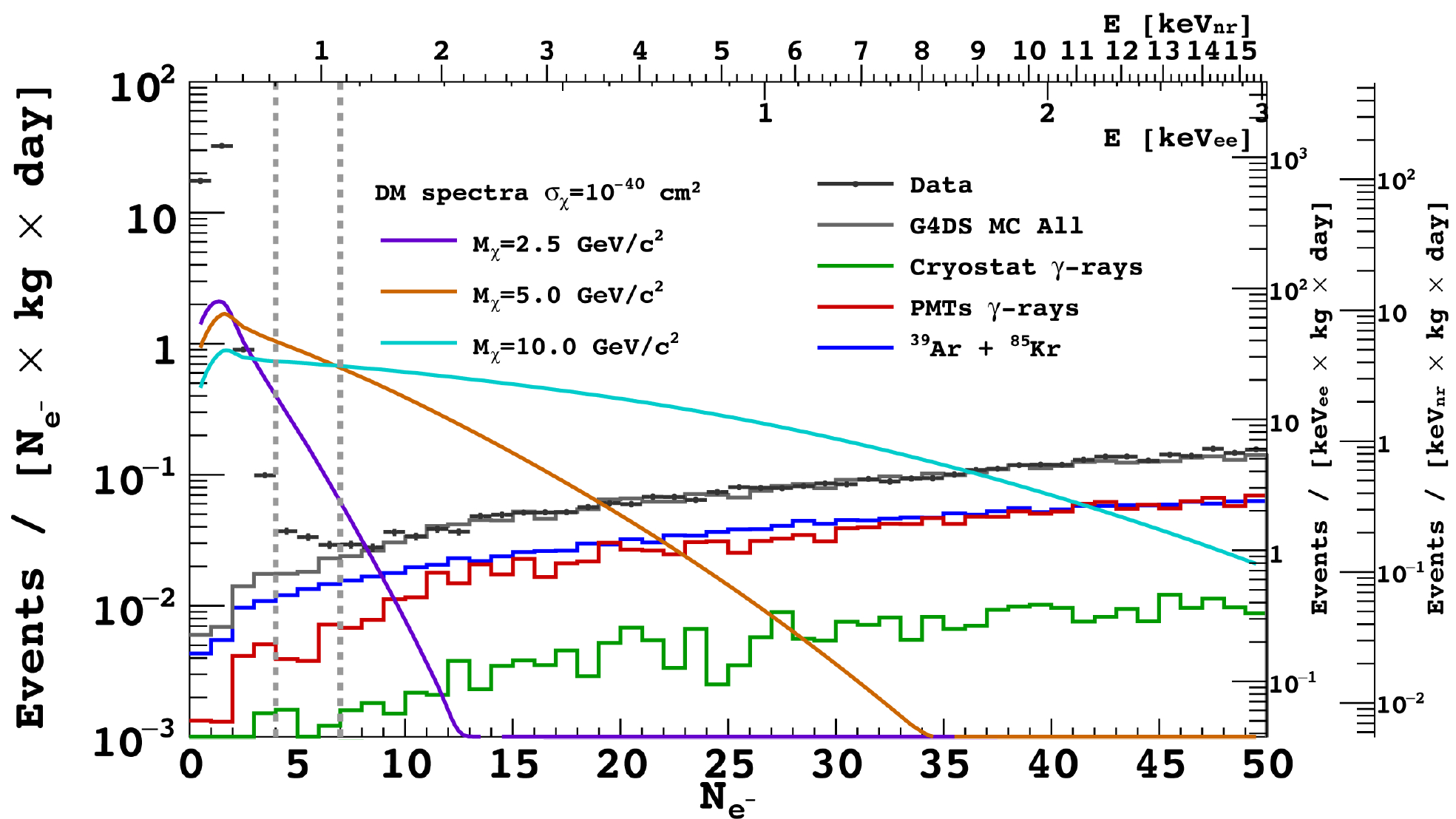}
		\caption{Fig.~\ref{fig_3.a}. The S2O analysis on DarkSide-50 6786.0 kg*d exposure shown the consistency of the expected background (the ``G4DS MC All'' curve) and observed events (the ``Data'' histogram). The plot is copied from reference~\cite{DarkSide502018S2O}.}\label{fig_3.a}	\end{subfigure}
	\quad
	~~~\begin{subfigure}[t]{3.0in}
		~~~\centering
		~~~\includegraphics[scale=0.55, angle = 0]{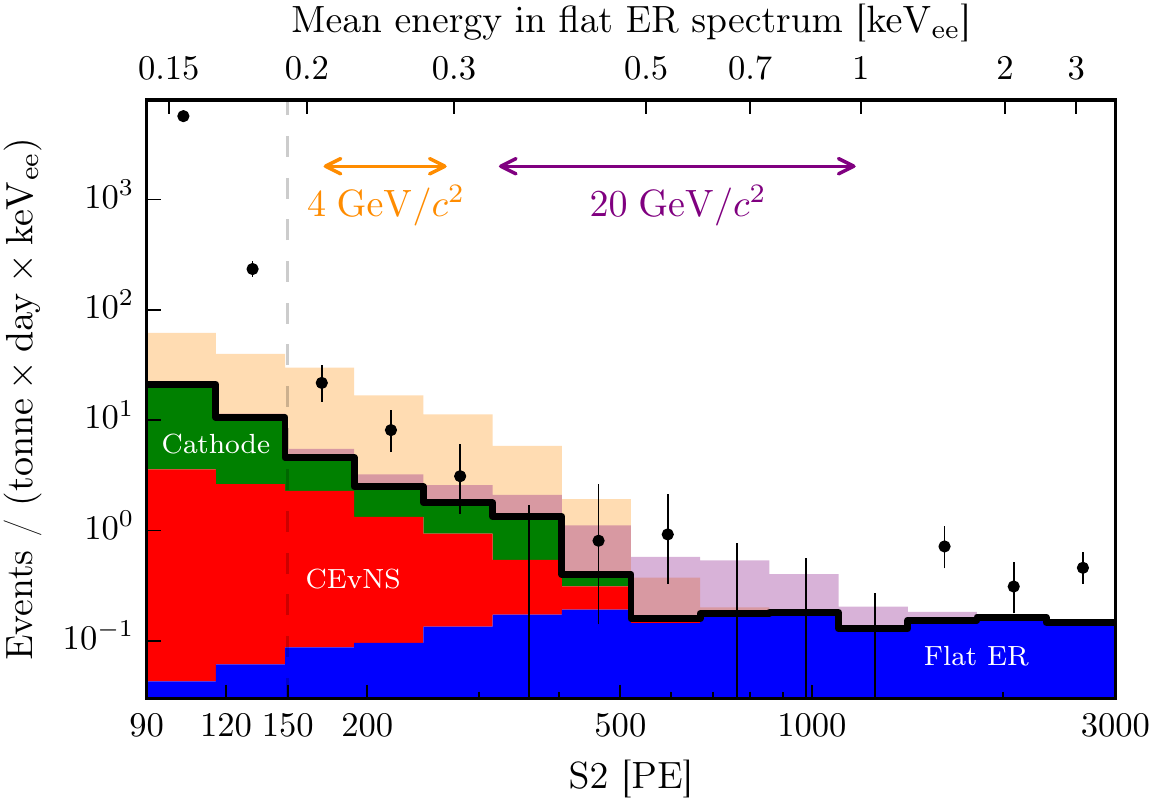}
		~~~\caption{Fig.~\ref{fig_3.b}. The S2O analysis on DarkSide-1T 22 $\pm$ 3 ton*day exposure. The expected background is shown as the thick black line. The observed events are black dots. The plot is copied from reference~\cite{XENON1TS2O19}.}\label{fig_3.b}
	\end{subfigure}
	\caption{The selected events after all of the cuts VS simulated background model for DarkSide-50~\ref{fig_3.a} and XENON-1T~\ref{fig_3.b}.}\label{fig_3}
\FloatBarrier
\end{figure}

Moreover, DarkSide-50 has successfully explored the ``S2 Only'' (S2O) channel for the WIMPs search, which extends the exclusion region for dark matter below previous limits~\cite{DarkSide502018} in the range of 1.8 GeV/c$^2$ - 6.0 GeV/c$^2$ WIMPs~\cite{DarkSide502018S2O}. Different from the PSD and ``S1/S2'' analysis, the S2O utilized the ``S2'' signal or ionization signal only (did not rely on S1 signals or scintillation.) to see if there exist more ER events than expected from backgrounds. One of physical motivations for S2O analysis was DM-electrons scattering~\cite{Essig12}.  The critical issue for such an analysis is to choose an appropriate fiducial volume where the observed events are consistent with expected background events. Fig.~\ref{fig_3.a} shows for S2 signals greater than seven e$^-$, the observed events are consistent with background events.

\captionsetup[subfigure]{labelformat=empty}

\section*{The legacy of  LXe TPC experiments in DM direct detection}

Among stable noble gases, xenon is the heaviest one, which is the most suitable element for the hunting of  high-mass WIMPs ($\mathcal{O}$ (100) GeV/c$^2$) due to the hypothesized coherent scattering between WIMPs and nucleons. As a result, the cross-section follows the ``$\text{A}^2$ law'' where ``A'' is the atomic number of the material of a detector. The bigger the $A$ value, the greater the cross-section. Since 2000, LXe TPCs have been widely applied for  high-mass WIMPs search in a few collaborations: ZEPLIN experiments~\cite{ZEPLINPubWeb}, XENON experiments~\cite{XENONPubWeb}, LUX~\cite{LUXPubWeb}, XMASS~\cite{XMASSWeb}, and PandaX~\cite{PandaXPubWeb}. These successful experiments have set the lowest limits on the high-mass WIMPs region one after another. As the biggest LXe TPC under commission, with 16.8 ton*years exposure, LZ is supposed to have only $\sim$ 6.2 ER and $\sim$0.6 NR background events (ER backgrounds could be further mitigated with improved removal of $^{85}$Kr and $^{222}$Rn.), achieve the lowest cross-section of WIMPs-nucleon for SI model down to $\sim 2.3$ 10$^{-48}$ cm$^2$ for 40 GeV/c$^2$~\cite{LZTDR17}.

LXe TPCs have successfully demonstrated to be capable of: (1) discriminating background-like ER events from signal-like NR events with the ``S1/S2'' analysis method and, (2) creating a fiducial volume where extremely low background events can be achieved and, (3) running stably for years and taking a leading role in search of  high-mass WIMPs. 
  
Similar to LAr experiments such as DarkSide-50, XENON collaboration has applied the S2O analysis into a series of LXe detectors, XENON-10~\cite{XENON10S2O11,XENON10S2O11Erratum}, XENON-100~\cite{XENON100S2O16, XENON100S2O16Erratum}, and XENON-1T~\cite{XENON1TS2O19}. Fig~\ref{fig_3.b} shows expected backgrounds (thick black line) and measured backgrounds (black dots) in XENON-1T.  

In summary, with a review on the legacy of LAr and LXe TPC experiments, we could conclude that the TPC technology has a few significant advantages such as: (a), providing a fiducial volume (X, Y and Z dimensions) with extremely low or zero background inside, which makes these experiments be very competitive; (b), discriminating ER/NR events with reliable analysis method, either PSD, or S1/S2, or both; (c), implementing an additional analysis method for WIMPs search: S2O; (d) recycling and purifying liquid noble gas to achieve very high purification; (e), calibrating the system online regularly to understand the performance of the detector; (f), scaling up from a smaller size to bigger size viable; (g), transplanting from one experiment to another one (for instance, XENON-10 to LUX, LUX to LZ, etc.), from one liquid noble gas to another (for instance, xenon to argon). 

We are transplanting the TPC technology into the low-mass sector by filling it with liquid helium (temperature is around 4 K, not in a superfluid state.). We were hoping all these technical advantages demonstrated in LAr and LXe TPCs could also be implemented in a liquid helium TPC.

\section*{The design of the ALETHEIA project}
The HERON (HElium Roton Observation of Neutrinos) project aims at creating a superfluid helium detector to detect Solar neutrinos~\cite{HOERN1}. The researches also considered using it for dark matter hunting~\cite{HOERN88, HOERN13}. Different from these earlier concepts, the ALETHEIA project will use liquid helium at $\sim$ 4 K, which is above the Lambda curve, instead of superfluid helium. Reference~\cite{GuoMckinsey13} also proposed and simulated a liquid helium TPC hunting for DM in 2013 with ``S1/S2'' analysis, where ``S1'' refers to prompt scintillation while ``S2'' is the electroluminescence light originated from an ionization by recoiled target nuclei. Besides the ``S1/S2'' analysis, the ALETHEIA  will also implement the S2O analysis to reach the sensitive WIMPS mass all the way down to 10s MeV/c$^2$~\footnote{The senGuoMckinsey13sitive mass region of the S2O analysis strongly depends on the data, as discussed below. However, according to the analysis of DarkSide-50~\cite{DarkSide502018S2O} and XENON-1T~\cite{XENON1TS2O19}, the S2O analysis can extend its sensitive mass roughly one order lower than the mass with traditional methods of PSD or S1/S2, from $\sim$ 30 GeV/c$^2$ to $\sim$ 2 GeV/c$^2$.}.  

\begin{figure}[!t]	 .
	\centering
        \includegraphics[width=5.0in, angle = 0]{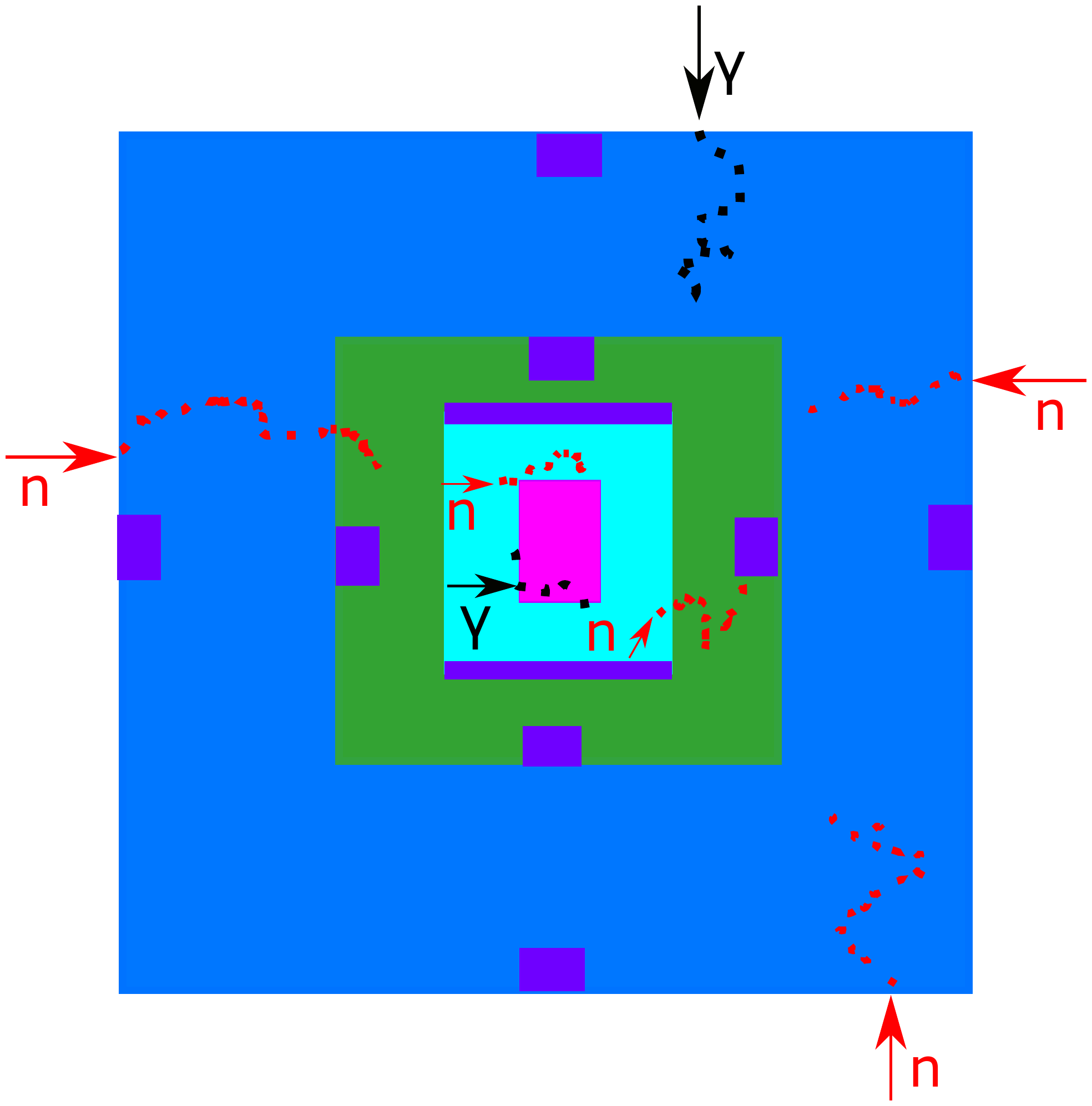}   
	\caption{The schematic drawing of the ALETHEIA detector (not to scale). From outside to inside: The light blue area represents the water tank surrounding the whole detector system, with a diameter of a few meters; the four purple rectangles on the edge of the water tank are the PMTs to detect background signals insides of the water tank; the dark green is the Gd-doped liquid scintillator veto, with thickness of $\sim$ half-meter; the four purple rectangles at the edge of veto (green area) are PMTs to detect signals insides of it; the cyan area is the active volume of the TPC, filled with liquid helium; the two horizontal dark blue stripes on the top and bottom of the active volume represent the SiPMs to detect scintillation and electroluminescence; the pink region represents the fiducial volume of the TPC where extremely low backgrounds are expected there. Red dots represent background neutrons. Black dots are  background $\gamma$s.}
	\label{FigSchematicALETHEIA}
\FloatBarrier
\end{figure}

The design of the dual-phase ALETHEIA has been significantly inspired by other successful experiments like DarkSide-50, LUX/LZ, PandaX, XENON-100, and XENON-1T, and others. The photon sensors would be SiPMs to achieve the highest possible photon detection efficiency. Figure~\ref{FigSchematicALETHEIA} shows the schematic drawing of such a detector.  

As shown in Fig.~\ref{FigSchematicALETHEIA}, the core of the ALETHEIA experiment is a dual-phase liquid helium TPC (in cyan). The TPC center is in pink representing the fiducial volume where extremely low or zero background is expected. On the top and bottom of the TPC are SiPMs (purple). The TPC was surrounded by a Gd-doped scintillator detector, which acts as a veto. The outmost is a water tank with a diameter of a few meters to shield neutrons and gammas outside of the detector system. 

For neutrons that come from outside of the water tank, a few meters of water should thick enough to thermalize them. The $\sim$ half-meter Gd-doped liquid scintillator would capture thermalized neutrons. For neutrons originated from the TPC inside, it could be identified by the feature of multiple hits in the TPC and (or) liquid scintillator. Typical WIMPs signal would only have one hit registered due to a much lower coupling constant between WIMPs and helium nuclei than the strong interaction between a neutron and helium nuclei. For $\gamma$s from outside of the water, the water tank can block them from entering the central detector; for $\gamma$s from inside of the TPC, ``S1/S2'', or PSD (Pulse Shape Discrimination), or a hybrid analysis combined both could discriminate from nuclear recoils induced by a neutron or a WIMPs.

\section*{Why $^4$He}

$^4$He is suitable for  low-mass WIMPs search thanks to several advantages: 

(1), High recoil energy. The same kinetic energy of incident WIMPs would result in greater recoil energy than any other heavier elements. Hydrogen is even lighter, but the quenching factor of Hydrogen is more than one order smaller than Helium at the recoil energy of $\sim$1 keV$_{\text{nr}}$. For details, please refer to the following (2).

(2), The Quenching Factor (QF) of LHe is quite high. For instance, for a 16 keV nuclear recoil, the measured QF of LHe is $\sim$ 65\%~\cite{Santos08}; while LAr is $\sim$ 24\%~\cite{QFArgon14}, which is a factor of 3 smaller. The measured Quenching Factor (QF) of Helium at 1.5 keV recoil energy is up to 22\%~\cite{Santos08}. As a comparison, the measured QF of Hydrogen at 100 keV is only 2\%~\cite{Reichhart12}, and the estimated QF at 1.5 keV nuclear recoils would be much lower. As a result, the QF of Hydrogen is guaranteed to be at least one order smaller than $^4$He. Thus, Hydrogen is not an appropriate material for  low-mass WIMPs search with the method of ionization, while Helium is.

(3), $^4$He only has 4 electrons. Therefore the intrinsic ER background induced by (background) gammas would be significantly smaller than other widely used heavier noble elements like Argon and Xenon. 

(4), At 4 K, only $^3$He is solvable in LHe, and $^3$He is very few in nature; other elements become a solid state at this low temperature. So, it would be easy to purify LHe with getter and cold trap technologies~\cite{HeliumPurificationJLab} to achieve very high level purification.  

(5), LHe is relatively cheaper. The price of LHe is $\sim$ 1/7 of LXe. ($^3$He is very expensive, which is partially the reason why it has been ruled out for consideration as a material for  low-mass WIMPs search.).

\section*{The review of the ALETHEIA project}

In Oct 2019, we organized a DM workshop at Peking University in Beijing, China~\cite{DMWS2019PKU}. Dr. J. Liao presented the project's concept during the workshop~\cite{LiaoPresentationDMWS2019PKU}~\footnote{The then name of the project is ALHET which stands for A Liquid HElium Time projection chamber.} and provided a $\sim$20-page documents (main text)~\cite{ALETHEIA_2019WorkPlan} to address the project in details. A panel composed of leading physicists~\footnote{The review panel member are: Prof. Rick Gaitskell at Brown, Prof. Dan Hooper at Fermilab at the University of Chicago, Dr. Jia Liu at the University of Chicago (now an assistant Prof. at Peking University.), Prof. Dan McKinsey at UC Berkeley, Dr. Takeyaso Ito at Los Alamos National Laboratory, and Prof. George Seidel at Brown University.} in DM and liquid helium reviewed the ALETHEIA project~\footnote{ALHET is the then name of the project. It's current name is ALETHEIA.}. The panel stated ``It is possible that liquid helium (TPC) could enable especially low backgrounds because of its powerful combination of intrinsically low radioactivity, ease of purification, and charge/light discrimination capability''~\cite{ALETHEIA_2019Review}. 
The panel suggested two R\&D phases: 30 g and 10 kg LHe detectors. 
The 30 g phase represents a cell having a total mass of 30 g LHe, with a cylindrical shape, radius = 5 cm, height = 3 cm. There will be multiple versions of this setup to test: \\
Cal-I: ER and NR calibrations, \\
Cal-II: S2 signal optimizations, \\
Cal-III: SiPM testing at 4 K. \\
The 30 g detector program is suited to answer the initial questions concerning the fundamental responses of the liquid helium to incident particles (neutrons and gammas/electrons) and also to establish the optimum conditions for operation. 

The subsequent 10 kg system ($\sim$ 10 kg LHe, cylindrical shape, diameter = height = 45 cm.) would be necessary in order to demonstrate the viability of the elevated HV levels needed for an even larger scale dark matter search experiment. It would also test other needed aspects such as large multi-channel photodetector arrays. Given the potentially slow drift speed of the ionization signals in a LHe TPC (Typical velocity of electron bubbles in LHe is $\sim$ 2 m/s.), to constrain the overlaps in background events due to  cosmic rays, the 10 kg detector would be operated underground, instead of running above ground.

As to the competitiveness of the project, the panel mentioned ``In the field of low-mass dark matter particle detectors (based on our current best estimates of the likely competing technologies) the full ALHET-10 kg  He target with a 4 keV threshold (defined as $>$50\% efficiency) operated underground in a low background environment would achieve very competitive science, if it started conducting a search sometime within 3-5 years.''~\cite{ALETHEIA_2019Review}. They didn't review the post 10 kg phase since ``their viability and design specifics are very contingent on the results from the earlier 30 g and 10 kg prototype phases''~\cite{ALETHEIA_2019Review}.

\section*{LHe TPC R\&D major questions and recommendations}
We outline here the questions to be demonstrated for a functional LHe TPC. Thankfully, the review panel addressed a list of R\&D activities and presented useful suggestions~\cite{ALETHEIA_2019Review}, much of which will be cited in the following. 

\subsection*{Electron Detection}
According to reference~\cite{SeidelPKUTalk2019}, it is possible that the PSD of LHe might not as powerful as LAr to discriminate ER from NR events. If this feature were confirmed experimentally, ``S1/S2'' or ``S1/S2'' + PSD analysis would instead be employed to do ER/NR discrimination. Consequently, electron detection would become the most critical and challenging part of the R\&D program because single-electron detection should be demonstrated. We have several approaches to achieve electron detection.\\
(a), Extracting electrons into GHe (Gaseous Helium), producing proportional scintillation light as other two-phase xenon and argon TPCs. \\
(b), GEM-based electron detection in the LHe or GHe. Reference~\cite{Erdal15} demonstrated this technology in an LXe apparatus, with a thick-GEM to multiply electrons in a GXe bubble produced by a heated wire beneath. \\
(c), Electroluminescence due to electrons entering very high fields surrounding thin wires immersed in the LHe. This method's mechanism is similar to a gas proportional tube where electrons avalanche happens as long as electrons approaching the anode of the tube.\\
(d), Charge amplification using small-scale structures, such as Micromegas.\\

\subsection*{Electron yields}
\begin{figure}	
	\centering
	\begin{subfigure}[t]{3.0in}
		\centering
		~~~\includegraphics[scale=0.3]{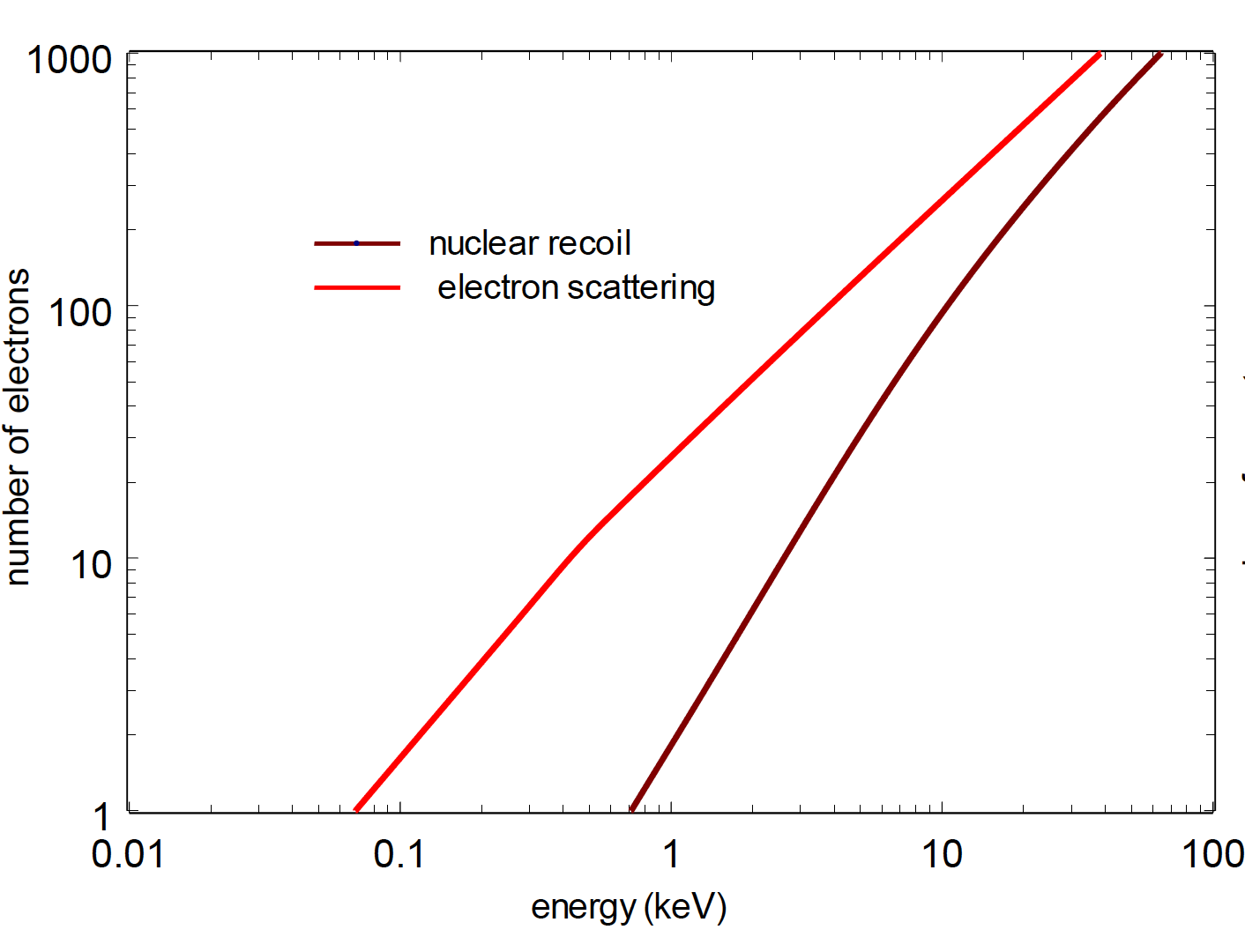}
		\caption{Fig.~\ref{fig_electrons.a}. The predicated electron yields. The plot is copied from reference~\cite{SeidelPKUTalk2019}.} \label{fig_electrons.a}	
	\end{subfigure}
	\quad
	~~~\begin{subfigure}[t]{3.0in}
		~~~\centering
		~~~\includegraphics[scale=0.3, angle = 0]{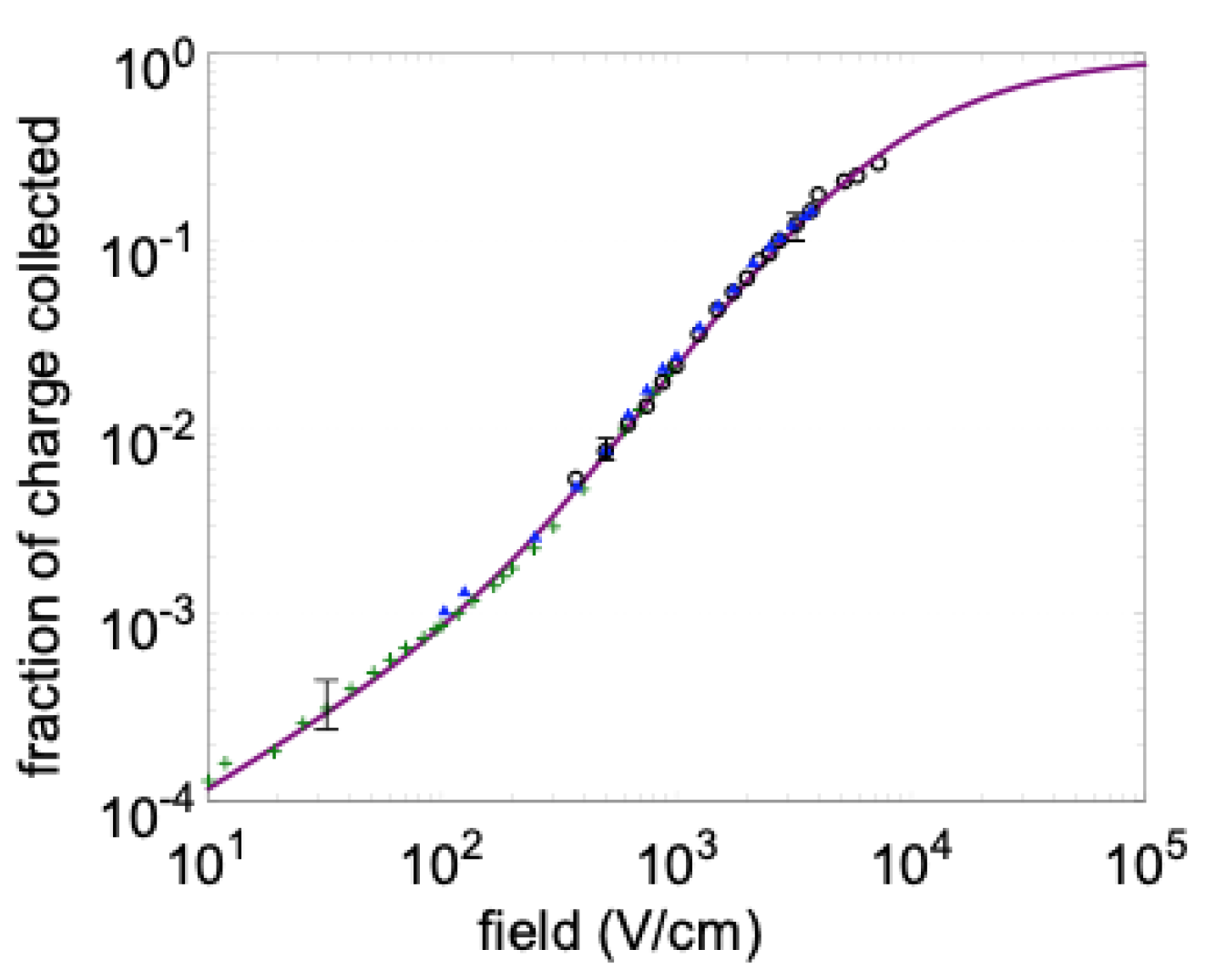}
		~~~\caption{Fig.~\ref{fig_electrons.b}. The fraction of electrons escaped from recombination for ER events at 2.5 K V.S. the external electric field.} \label{fig_electrons.b}
	\end{subfigure}
	\caption{ The electron production and collection in LHe.  }\label{fig_electrons}
\FloatBarrier
\end{figure}

The predicated electron yield is shown in Fig.~\ref{fig_electrons.a}, which is copied from reference~\cite{SeidelPKUTalk2019}. The prediction is based on the measured cross-section of He-He scattering and the assumed average energy to ionize a pair of electron-ion, 43 eV. The fraction of electrons escaped from recombination (for ER events) by the application of an external electric field at 2.5 K is shown in Fig.~\ref{fig_electrons.b} copied from reference~\cite{XenonLightYield10}. According to reference~\cite{XenonLightYield10},  the fraction is temperature dependent; the fraction at 4 K is around double to the one at 2.5 K. There is no direct measurement has been accomplished for NR events yet. We will measure electron yields in liquid helium for ER and NR at 4 K with one of our 30 g LHe apparatuses.

\captionsetup[subfigure]{labelformat=empty}

\subsection*{Drift velocity}
An electron in liquid helium forms a bubble of 19 {\AA} radius as long as its energy thermalize to $\sim$1 eV~\cite{Maris08}. Consequently, the electron bubble has a low mobility as $2 \cdot 10 ^{-2} cm^2/ (V \cdot s)$ at 4.2 K~\cite{Schwarz72}. In a field of $10^4$ V/cm, the velocity of an electron bubble is around 2 m/s. An electron will take around 15 ms to drift across the 3 cm height 30 g cell. For the cell with a radius of 5 cm, the horizontal surface is 80 cm$^2$. Since the cosmic ray flux at the sea level is around 1 / (cm $\cdot$ min)$^2$~\cite{PDG20}, which corresponds to 1.3  cosmic ray events per second in  the 30 g cell. During a 15 ms drifting time, the chance of observing a cosmic ray event is 2\%. This could be a viable test at a lab above ground. For the 10 kg LHe apparatus with diameter = height = 45 cm, the drifting time is about 0.25 s. The cosmic ray would register about 26 events in the detector per second, or 6.5 events per 0.25 seconds. That's being said, during the time interval of an electron bubble drifting in the 10 kg LHe cell, there will be 6.5 cosmic ray events hitting the same detector. As a result, the 10 kg detector should test at an underground lab, instead of a ground lab.

\subsection*{Scintillation yield}
There exist three scintillation components: prompt scintillation with a time constant of $<$ 10 ns resulted from the decay of directly excited singlets, He$^*_{2}( \text{A} ^{1} \sum_{\mu}^{+}) \rightarrow 2 \text{He}(1^{1} \text{S} ) + h\nu$; 1.6 $\mu$s originated from the decay of excited atoms $\text{He}^*(_2^1\text{S})$ formed singlet $\text{He}^*(_2^1\text{S}) \rightarrow $  He$^*_{2}( \text{A} ^{1} \sum_{\mu}^{+}) \rightarrow 2 \text{He}(1^{1} \text{S} ) + h\nu$ , and 13 s resulting from triplets decay~\cite{McKinsey03}. Because of the absence of information about the transition probabilities among the multitude of excited atomic and excimer states, predictions about the number of photons produced in the various channels are highly uncertain. Estimates of the yield range from a few tens of photons at 1 keV to several 100 at 10 keV NR energy~\cite{GuoMckinsey13, ItoSeidel13}. Observations at a few MeV NR energies show the $<$ 10 ns and 13 s components are comparable while the 1.6 $\mu$s emission is considerably weaker~\cite{McKinsey03}. However, for the interesting NR energy interval of the ALETHEIA, [ 0.5 keV$_{nr}$, 10 keV$_{nr}$], no measurement has performed to understand the feature of scintillation for ER and NR events yet. The ALETHEIA would launch such a dedicated calibration, which would determine whether the ALETHEIA could apply the PSD algorithm for ER/NR analysis or not (but won't affect the ``S1/S2'' analysis.).

The wavelength of the scintillation in LHe has a peak at 80 nm (or 16 eV) \cite{McKinsey03}, which is out of the detectable range of current commercially available photon sensors. So a wavelength shifter like tetraphenyl butadiene (TPB) is required to shift the photons to $\sim$450 nm~\cite{Lippincott12}. Reference~\cite{Benson18} measured the efficiency of TPB coating on an acrylic substrate at room temperature in a vacuum. They obtained $\sim$ 30\% wavelength shifter efficiency for 80 nm light. The reference also predicted a higher efficiency under LAr/LHe environment. 

The lower the WIMPs' mass, the lower the kinetic energy of WIMPs, the lower the ionizing energy, and the smaller the number of scintillation photons. As a result, we will choose SiPMs as our photon detectors, thanks to their higher detection efficiency. 
Given the small number of estimated EUV photons from low energy nuclear recoil events, the ALETHEIA will work with industrial companies to develop high photon-detection efficiency photosensors, which would hopefully be capable of operating at 4 K. 

Although no commercial companies tested the performances of SiPMs at LHe temperature, certain types of Hamamatsu SiPMs not designed to work at LHe temperature surprisingly turned out to be functional at such low temperatures as indicated in references~\cite{Cardini14, Iwai19}. Moreover, the relative PDE (Photon Detection Efficiency) near LHe temperature (5 K and 6.5 K) is roughly to be 70\% of room temperature according to the two independent tests.

\subsection*{Nuclear recoil calibrations}
Under a range of applied drift fields (up to 100 kV/cm), the ALETHEIA experiment would measure the experimental scintillation ($<$ 10 ns, 1.6 $\mu$s, and 13 s components) and ionization signal yields for interesting NR energy interval (0.5 - 10 keV$_{nr}$). Preliminary NR calibration would be completed with the 10$^5$ Bq PuBe neutron sources in our lab at CIAE. A more precise measurement is scheduled to implement the $\sim$ 50 - 100 keV mono-energy neutrons generated by a proton beam hitting on a $^7$Li target on a Tandem Van de Graaff accelerator also at CIAE, with the interaction of $^7$Li(p,n)$^7$Be.

\subsection*{High voltage in the ALETHEIA}
As mentioned above, an electric field equal to or larger than 10 kV/cm is needed to achieve sufficient charge detection. The LANL (Los Alamos National Laboratory) group has applied $>$ 100 kV/cm in a volume between 12 cm diameter electrodes separated by 1 cm~\cite{Ito16}, which indicates applying a 100-kV electrical potential into liquid helium from an external power supply component is viable. Introducing a 200 kV into liquid helium from an external commercial power supply is likely possible with custom-designed HV feed-though~\cite{Cantini17}. For higher potential, generating HV directly inside the liquid, using eg Cavallo's multiplier~\cite{Clayton18}, is likely necessary.

\section*{Cryogenics, calibrations, electronics and DAQ, recycling and purification, screening, and simulation}
At current stage, the discussions of cryogenics, calibrations, electronics and DAQ, recycling and purification, screening, and simulation are less urgent to be addressed than the topics already have discussed in the document. We would supply in next update.

\section*{Implementing PSD, S1/S2, and S2O into the ALETHEIA for WIMPs search}
\subsection*{PSD for ALETHEIA}
\begin{figure}[!t]	
	\centering
        \includegraphics[width=6.0in, angle = 0]{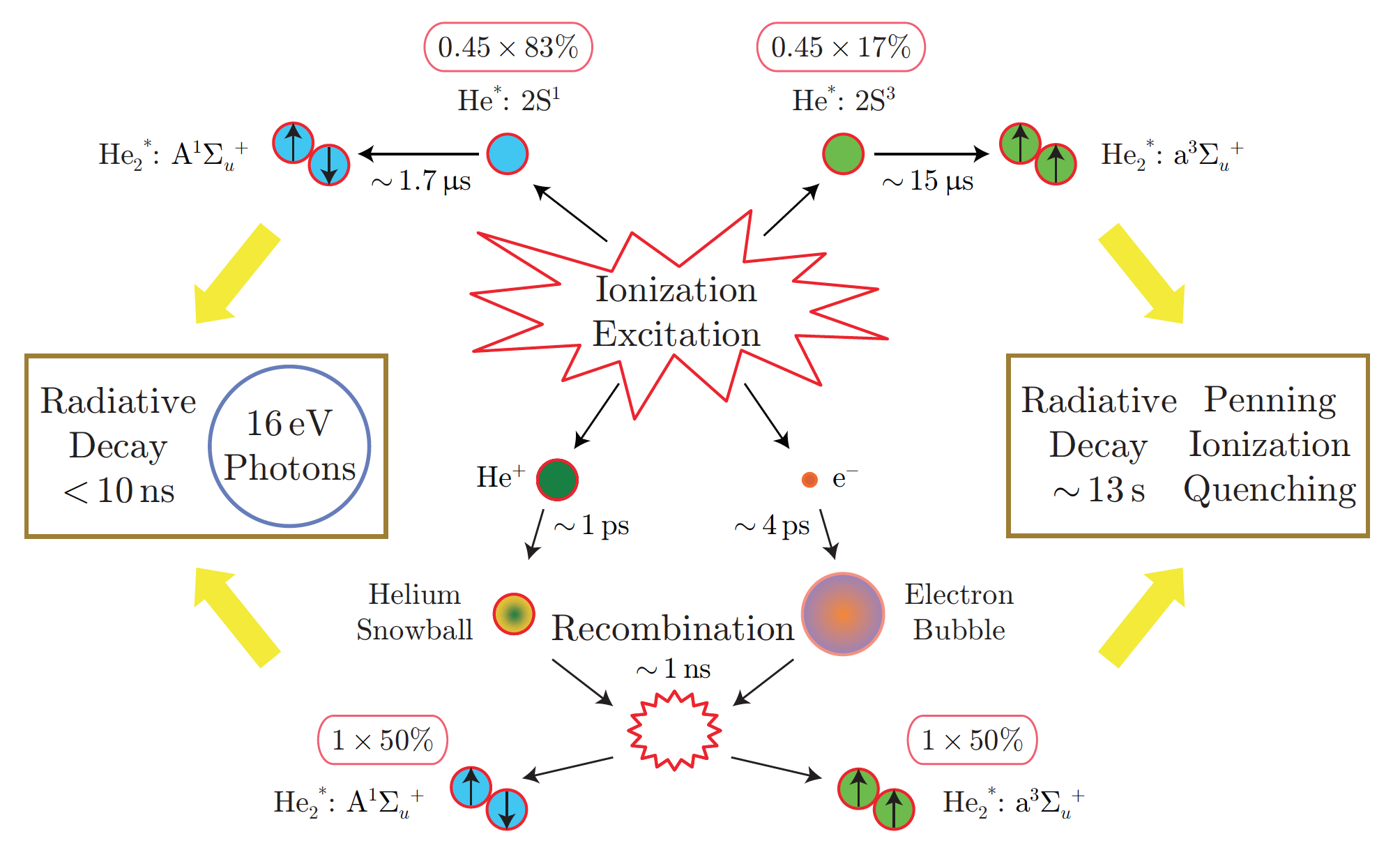}
	\caption{100 keV electrons induce excitation and ionization in liquid helium. The yield of scintillation was calculated (not measured) in reference~\cite{ERLHeCalculation74}. The plot is copied from reference~\cite{DaifeiJinPhDThesisBrown12}.}\label{FigScintillationGeneration}
\FloatBarrier
\end{figure}

Energetic particles passing through a medium of Liquid $^4$He (LHe)  will deposit part or all of its kinetic energy. If the incident particle is an electron (or $\gamma$), it interacts with the electrons of helium atoms electromagnetically. The atoms then get ionized or excited, or both. Fig.~\ref{FigScintillationGeneration} schematically shows the procedure of both processes with 100 keV incident electrons~\cite{DaifeiJinPhDThesisBrown12}. 

If the incident particle is a neutron, it interacts with helium atoms strongly. The atoms get recoiled energy and become moving $\alpha$ particles. The $\alpha$ particles further interact with the electrons of surrounding helium atoms electromagnetically. Although electrons and $\alpha$ particles both interact with helium atoms via electromagnetic interaction, the charge densities of ions and electrons are different. As mentioned in reference~\cite{McKinsey03}, for high-energy ($\sim$ MeV) electrons, energy deposition is 50 eV / $\mu$m$^{-1}$; while for $\alpha$ particles, energy deposition is 2.5 $\times$10$^4$ eV / $\mu$m$^{-1}$. For liquid xenon, similar results were obtained with simulation~\cite{DahlPhDThesis09}. 

\begin{figure}[!t]	 
	\centering
        \includegraphics[width=6.0in, angle = 0]{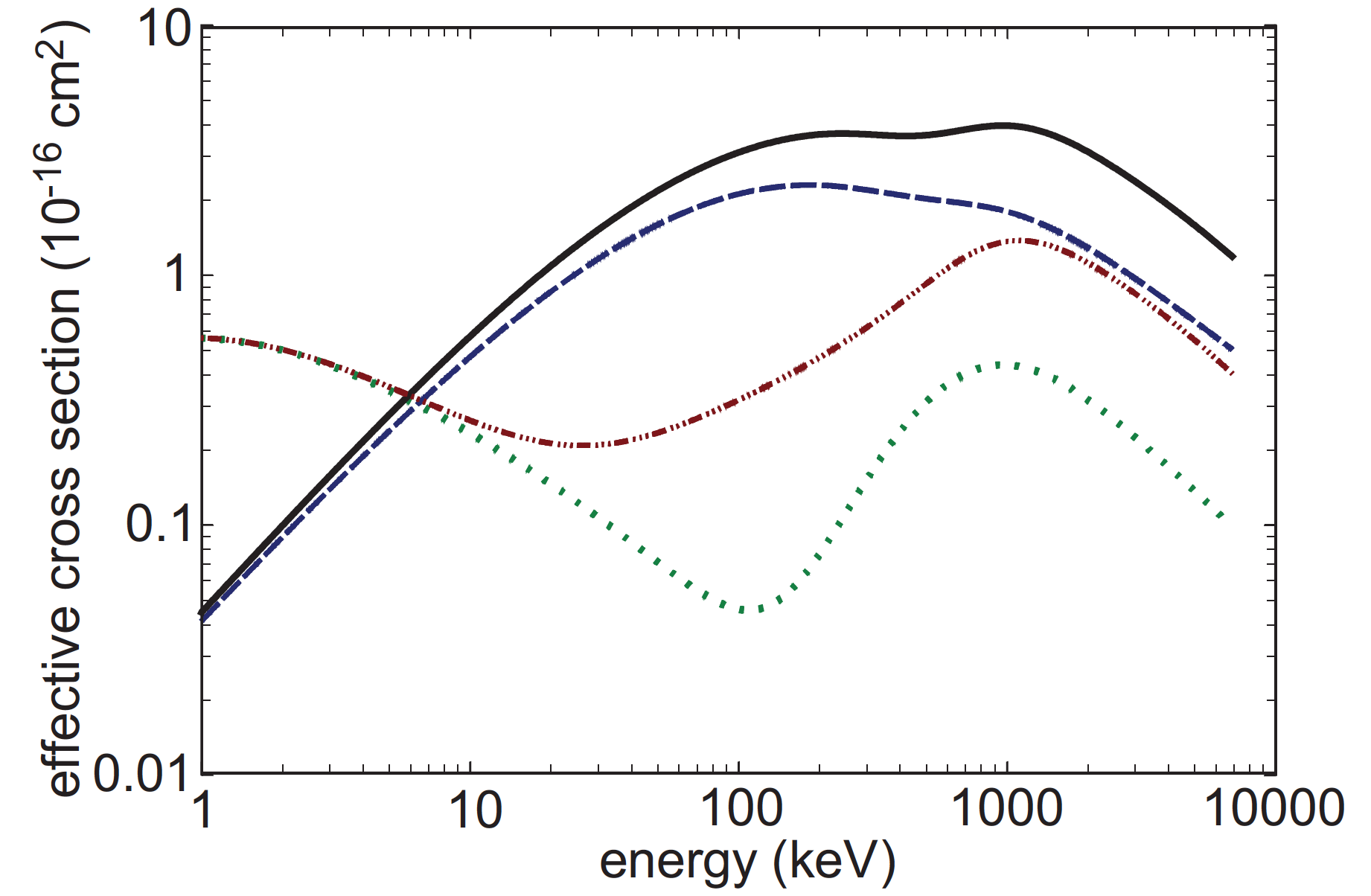}
	\caption{Effective cross-section of excitation and ionization for liquid helium hitting by $\alpha$ particles. The X-axis is the energy of incident $\alpha$ particles~\cite{ItoSeidel13}. The cross-section curve fits well on many variant experimental datasets (not shown on this plot). Solid line: Ionization with secondary electron contribution. Dashed line: Ionization without secondary electron contribution. Dot-dashed line: Excitation with secondary electron contribution. Dotted line: Excitation without secondary electron contribution. The plot is copied from~\cite{ItoSeidel13}.}\label{FigIonizationExcitationEffectiveXS}
\FloatBarrier
\end{figure}

The effective cross-sections of ionization and excitation is shown in Fig.~\ref{FigIonizationExcitationEffectiveXS}, which are copied from reference~\cite{ItoSeidel13}. 

Essentially, a charge density depends on stopping power, or ``dE/dx''~\cite{McKinsey03, AprileDokeReview2009}. According to the \textit{Bethe formula}, for low energy incident particles ( $v << c$, where $v$ is the velocity of particles, $c$ is the speed of light.), $dE/dx \sim \propto 1/v$, while for the same kinetic energy $\alpha$s and electrons (100 keV for instance), the velocity of electrons is roughly 3 orders faster than $\alpha$s, as a result, $dE/dx$ of electrons is roughly 3 orders smaller than $\alpha$s. Consequently, the charge densities for ER/NR have three orders difference. Moreover, the geometry and separation of the tracks induced by ER and NR are different: ER events are ``small dots'' shape and separated on average 500 nm; NR events are the cylindrical shape and separated, on average, 1 nm. Considering the distance between the ion and its separated electrons is roughly 20 nm, ER events are well separated (500 nm $>>$ 20 nm), the recombination is geminate~\cite{Onsager38}(meaning the recombined electron-ion pair is the one being separated by incident particles); while NR events are heavily overlapped between individual ionization events (1 nm $<<$ 20 nm)~\cite{McKinsey03}, the recombination is columnar~\cite{Jaffe1913, Kramers52}. Though to be verified experimentally for recoil energy in 0.5 - 10 keV$_{nr}$, the different ways of recombination might lead to different ratios of (fast components) / (slow component), which is the reason why the PSD is a possibly viable analysis channel for an LHe TPC.

The PSD technique exploits the time feature of scintillation to discriminate ER/NR events, specifically, the lifetime of scintillation produced in singlet and triplet excimers: For LAr, the lifetime of singlet and triplet are a few ns and $\mu$s, respectively. The two orders difference in a lifetime is proved to be able to apply PSD for ER/NR in DarkSide-50 and DEAP. For LXe, the difference of the lifetime for singlet and triplet is only one order. Therefore, LXe experiments have not implemented PSD.

LAr has a fast component scintillation, $\sim$ 7 ns, which decays from excited singlet; and a slow component of 1.6 $\mu$s, which decays from excited triplet~\cite{Lippincott08}. As mentioned above in the section of ``Scintillation yield'', LHe has a different scintillation feature, $< 10$ ns,  1.6 $\mu$s and 13 s~\cite{McKinsey03}.  According to the test with a few MeV incident $\alpha$ sources, the 1.6 $\mu$s component is weaker than the $< 10$ ns and 13 s~\cite{McKinsey03}. For the ROI energy of the ALTHEIA, 0.5 - 10 keV$_{nr}$, the relative light strength might be different since the effective cross-sections for excitation and ionization channels are recoil energy dependent; and the scintillation components yield depend on the channel, as shown in Fig.~\ref{FigScintillationGeneration}. For a few MeV recoil energies, the effective cross-section of ionization is around a factor of 5 greater than excitation; while for 0.5 - 10 keV$_{nr}$, the relative cross-section is flipped: the excitation is a factor of a few greater than ionization, as shown in Fig~\ref{FigIonizationExcitationEffectiveXS}. However, as mentioned in reference~\cite{SeidelPKUTalk2019}, predictions about scintillation light yield in the various channels are uncertain. Only dedicated experiments can conclude the viability of the PSD analysis on an LHe TPC.

\begin{figure}	
	\centering
	\begin{subfigure}[t]{3.0in}
		\centering
		\includegraphics[scale=0.22]{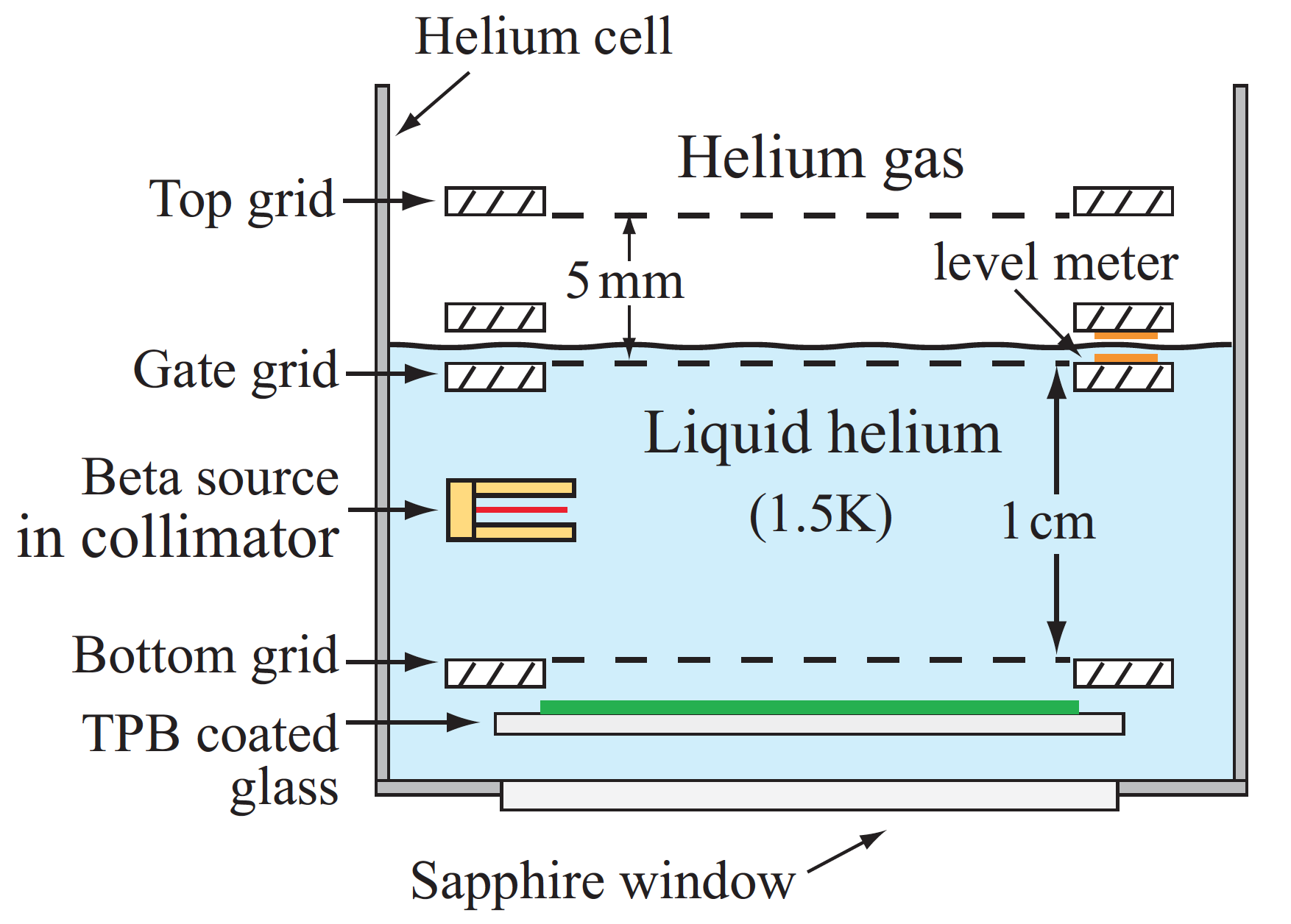}
		\caption{Fig.~\ref{fig_ERScintillation.a}. The 1.0 cm scale apparatus to do ER scintillation tests with a $^{90}$Sr source.}\label{fig_ERScintillation.a}	
		\end{subfigure}
	\quad
	\begin{subfigure}[t]{3.0in}
		\centering
		\includegraphics[scale=0.225, angle = 0]{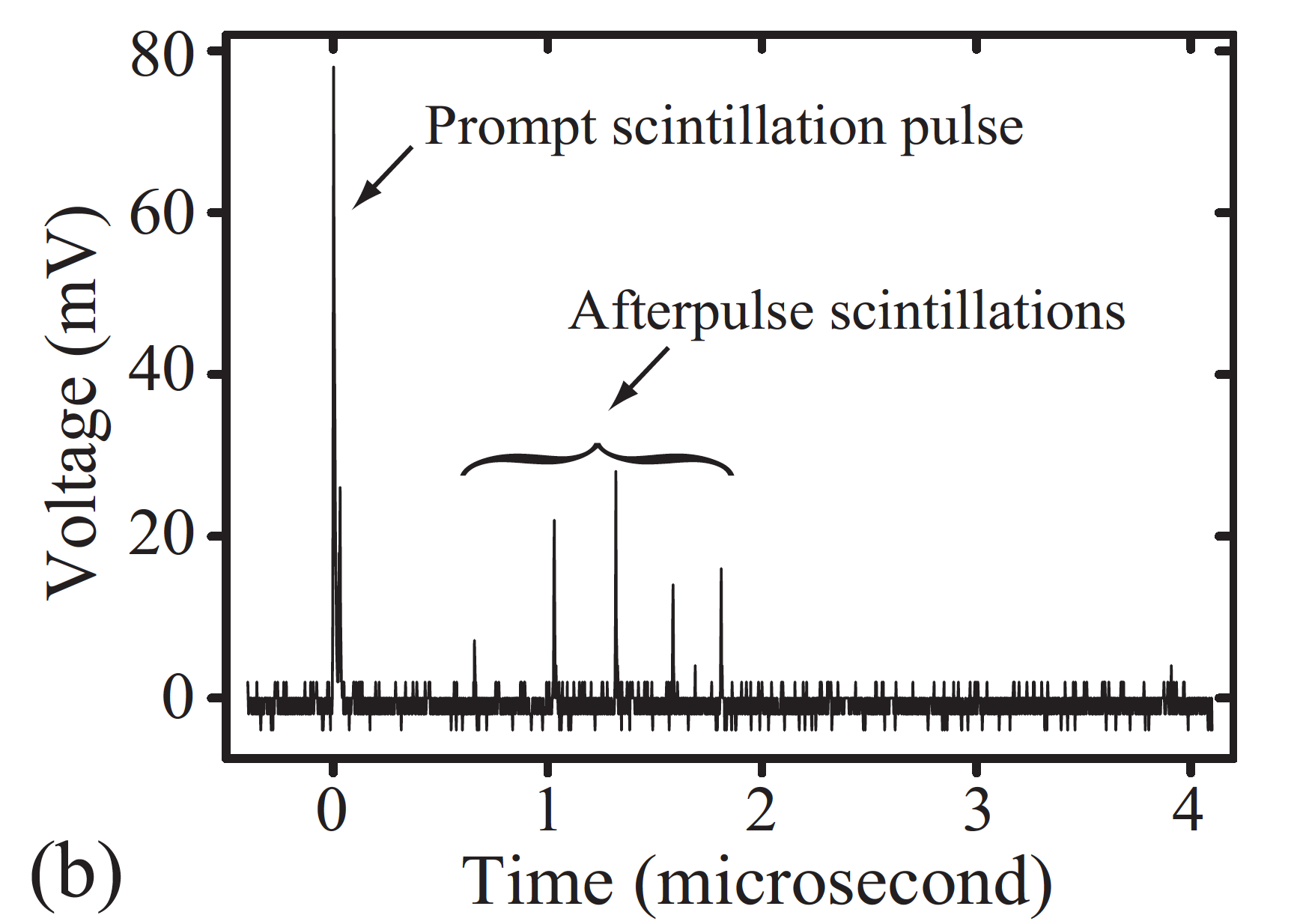}
		\caption{Fig.~\ref{fig_ERScintillation.b}.  Typical signals for a beta emission event in the liquid helium apparatus were observed and shown in Fig.~\ref{fig_ERScintillation.a}.}\label{fig_ERScintillation.b}
	\end{subfigure}
	\caption{The tests on ER scintillation with a $^{90}$Sr source. Both Fig.~\ref{fig_ERScintillation.a} and Fig.~\ref{fig_ERScintillation.b} are copied from reference~\cite{GuoJinst12}.}\label{fig_ERScintillation}
\FloatBarrier
\end{figure}

For ER events, with a 1.0 cm scale apparatus as shown in Fig.~\ref{fig_ERScintillation.a}, singlets scintillation (``prompt scintillation pulse'' in Fig.~\ref{fig_ERScintillation.b}) and triplets scintillation (``Afterpulse scintillations'' in Fig.~\ref{fig_ERScintillation.b}) in the fast components of LHe were explicitly observed~\cite{GuoJinst12}. The beta source is $^{90}$Sr, which has a decay energy of 0.546 MeV and a half-life of 28.8 years. The decayed yttrium isotope, $^{90}$Y, also undergoes $\beta^-$ decay and has decay energy of 2.28 MeV and the half-life of 64 hours.

\captionsetup[subfigure]{labelformat=empty}

\begin{figure}	
	\centering
	\begin{subfigure}[t]{3.0in}
		\centering
		\includegraphics[scale=0.21]{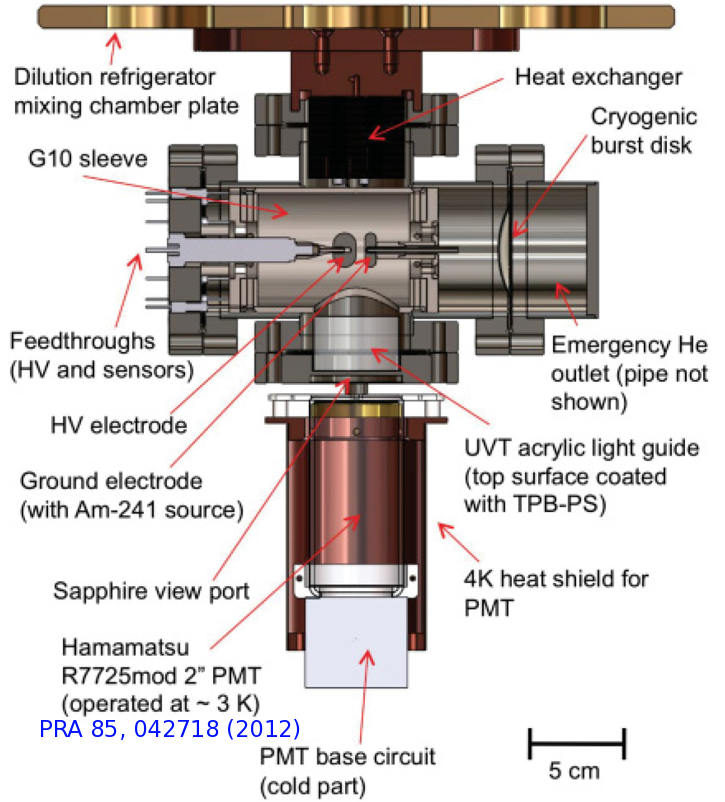}
		\caption{Fig.~\ref{fig_NRScintillation.a}. The 5.0 cm scale apparatus to study NR scintillation with an $^{241}$Am source.}\label{fig_NRScintillation.a}	
		\end{subfigure}
	\quad
	\begin{subfigure}[t]{3.0in}
		\centering
		\includegraphics[scale=0.3, angle = 0]{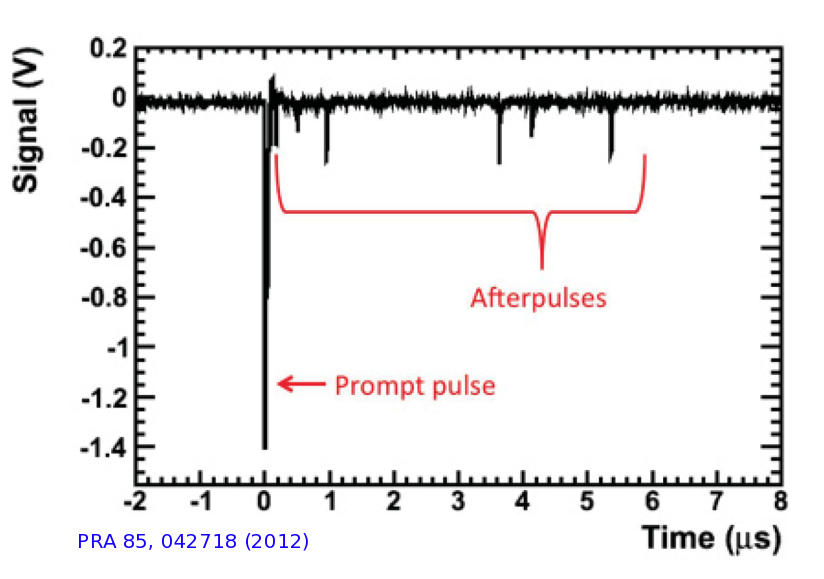}
		\caption{Fig.~\ref{fig_NRScintillation.b}.   Observed typical signals for $\alpha$ particle events in the liquid helium apparatus shown in Fig.~\ref{fig_NRScintillation.a}.}\label{fig_NRScintillation.b}
	\end{subfigure}
	\caption{The tests on NR scintillation with an $^{241}$Am source. Both Fig.~\ref{fig_NRScintillation.a} and Fig.~\ref{fig_NRScintillation.b} are copied from reference~\cite{Ito2012}.}\label{fig_NRScintillation}
\FloatBarrier
\end{figure}

Reference~\cite{Ito2012} studied NR events with a 5 cm scale apparatus as shown in Fig.~\ref{fig_NRScintillation.a}. An $^{241}$Am source with decayed 5.5 MeV $\alpha$ particles were used for the tests. Typical singlets scintillation (``prompt scintillation pulse'' in Fig.~\ref{fig_NRScintillation.b}) and triplets scintillation (``Afterpulse scintillations'' in Fig.~\ref{fig_NRScintillation.b}) in the fast components of LHe were explicitly observed~\cite{Ito2012}. 
\captionsetup[subfigure]{labelformat=empty}

\begin{figure}	
	\centering
	\begin{subfigure}[t]{3.0in}
		\centering
		\includegraphics[scale=0.45]{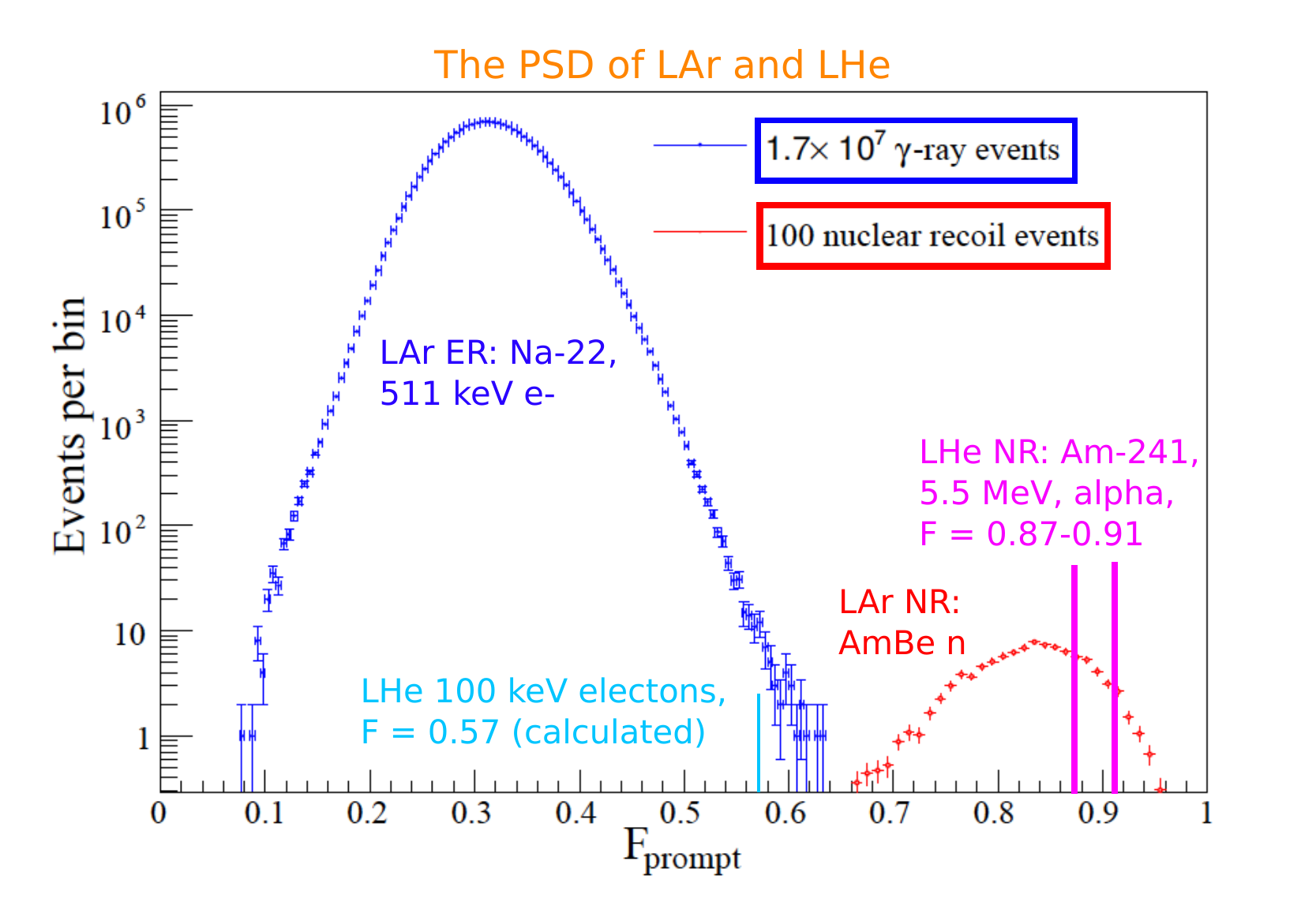}
		\caption{Fig.~\ref{fig_LHePSD_LArPSD.a}. The powerful PSD of LAr: 1.7 $\times 10^7$ ER events and 100 NR events were well separated~\cite{LArPSDDEAP09}. The ER of LHe was projected from Fig.~\ref{FigScintillationGeneration} and is shown as a vertical line where F$_{prompt}$ = 0.57. The yield of scintillation was calculated (not measured) in reference~\cite{ERLHeCalculation74}. The NR of LHe was projected from Fig.~\ref{fig_LHePSD_LArPSD.b} and is shown as the range of 0.87 $<<$ F$_{prompt} << 0.91$.}\label{fig_LHePSD_LArPSD.a}	
		\end{subfigure}
	\quad
	\begin{subfigure}[t]{3.0in}
		\centering
		\includegraphics[scale=0.26]{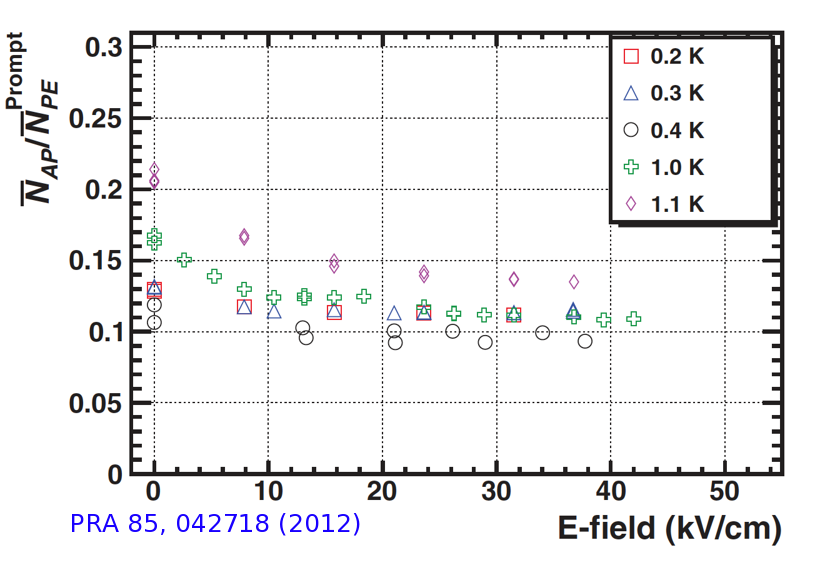}
		\caption{Fig.~\ref{fig_LHePSD_LArPSD.b}.  Measured ratio of $\bar{N}_{AP}$/$\bar{N}^{Prompt}_{PE}$. Projecting the 0.1 $ < \bar{N}_{AP}$/$\bar{N}^{Prompt}_{PE} < 0.15$ to F$_{prompt}$ would result 0.87 $<<$ F$_{prompt} << 0.91$, as shown in Fig.~\ref{fig_LHePSD_LArPSD.a}. The plot was copied from reference~\cite{Ito2012}.}\label{fig_LHePSD_LArPSD.b}
	\end{subfigure}
	\caption{The PSD of LAr and LHe.}\label{fig_LHePSD_LArPSD}
\FloatBarrier
\end{figure}

Fig.~\ref{fig_LHePSD_LArPSD.a} shows the powerful capability of ER/NR discrimination of LAr as demonstrated by the DEAP experiment~\cite{LArPSDDEAP09}: 1.7 $\times 10^7$ ER events and 100 NR events were well separated. Fig.~\ref{fig_LHePSD_LArPSD.b} is the measured ratio of $\bar{N}_{AP}$/$\bar{N}^{Prompt}_{PE}$, where $\bar{N}_{AP}$ is the number of after pulses, $\bar{N}^{Prompt}_{PE}$ is the number of prompt pulses. As shown in this figure, $0.1 \le \bar{N}_{AP}$/$\bar{N}^{Prompt}_{PE} \le 0.15$. If we change the notation to be consistent with LAr tests, F$_{prompt}$ = $\bar{N}^{Prompt}_{PE}$/ ($\bar{N}^{Prompt}_{PE}$ + $\bar{N}_{AP}$), the corresponding F$_{prompt}$ would be 0.87 $\le$ F$_{prompt} \le 0.91$, as the two pink vertical lines shown in Fig.~\ref{fig_LHePSD_LArPSD.a}. So, for 5.5 MeV $\alpha$ particles impinge LHe, the F$_{prompt}$ line up with the NR events of LAr. For 100 keV electrons, as shown in reference~\cite{ERLHeCalculation74}, the calculated yield of fast scintillation (10 ns) and slow scintillation (1.7 $\mu$s) are 1 $\times$0.5 = 0.5, and 0.45 $\times0.83$ = 0.3735, respectively. Using the formula of F$_{prompt}$ = $\bar{N}^{Prompt}_{PE}$/ ($\bar{N}^{Prompt}_{PE}$ + $\bar{N}_{AP}$), we get F$_{prompt}$ = 0.3735 / (0.3735 + 0.5) = 0.57, as the cyan vertical line shown in Fig.~\ref{fig_LHePSD_LArPSD.a}, which is consistent with the LAr tests.  In summary, with the only available data, calculated 100 keV electrons and measured 5.5 MeV $\alpha$ particles, the F$_{prompt}$ is consistent with the LAr results, which demonstrate that we can implement the PSD analysis to discriminate ER from NR events in ALETHEIA. Dedicatedly designed calibrations should be performed to see if the PSD is applicable for LHe in $\sim$ 0.5 keV$_{nr}$ $<$ E$_{nr} < $ 10 keV$_{nr}$.

\captionsetup[subfigure]{labelformat=empty}

\subsection*{S1/S2 for ALETHEIA}

LXe dark matter experiments have demonstrated that the S1/S2 technique can be utilized for discriminating ER/NR events and defining a fiducial volume (with S2). In principle, the same analysis methods could be transplanted into the ALETHEIA because the difference of charge density of ER and NR events in LXe also exists in LHe~\cite{McKinsey03}. 

\begin{figure}	
	\centering
	\begin{subfigure}[t]{3.0in}
		\centering
		\includegraphics[scale=0.29]{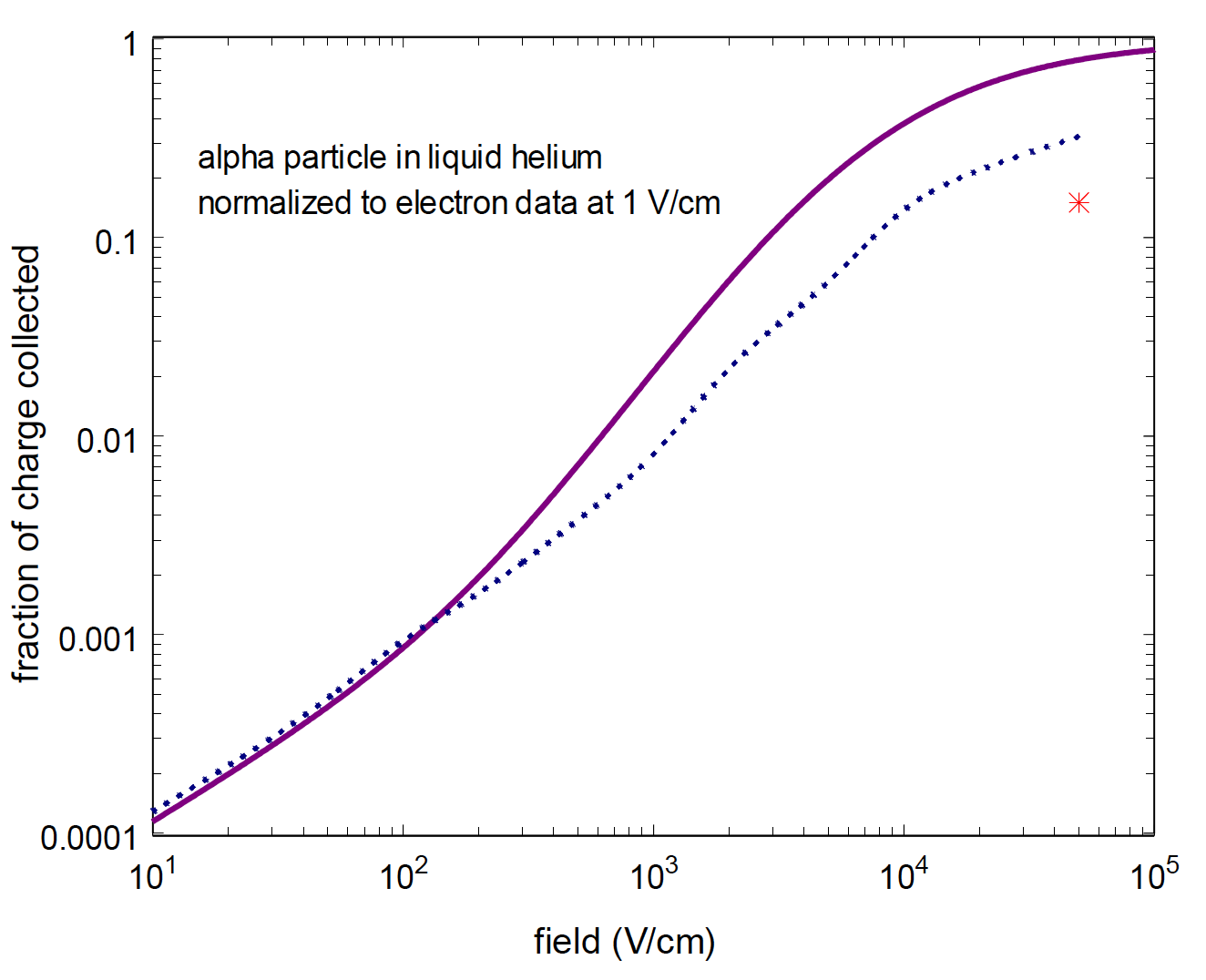}
		\caption{Fig.~\ref{fig_LHeS1S2.a}. Fraction of electrons extracted in LHe by $\alpha$ and $\beta$ particles. Solid line: fitted curve on $^{63}$Ni data. Doted line: calculated for $\alpha$ particles. Fig.~\ref{fig_LHeS1S2.a} was copied from reference~\cite{Seidel14}.}\label{fig_LHeS1S2.a}	
		\end{subfigure}
	\quad
	\begin{subfigure}[t]{3.0in}
		\centering
		\includegraphics[scale=0.2, angle = 0]{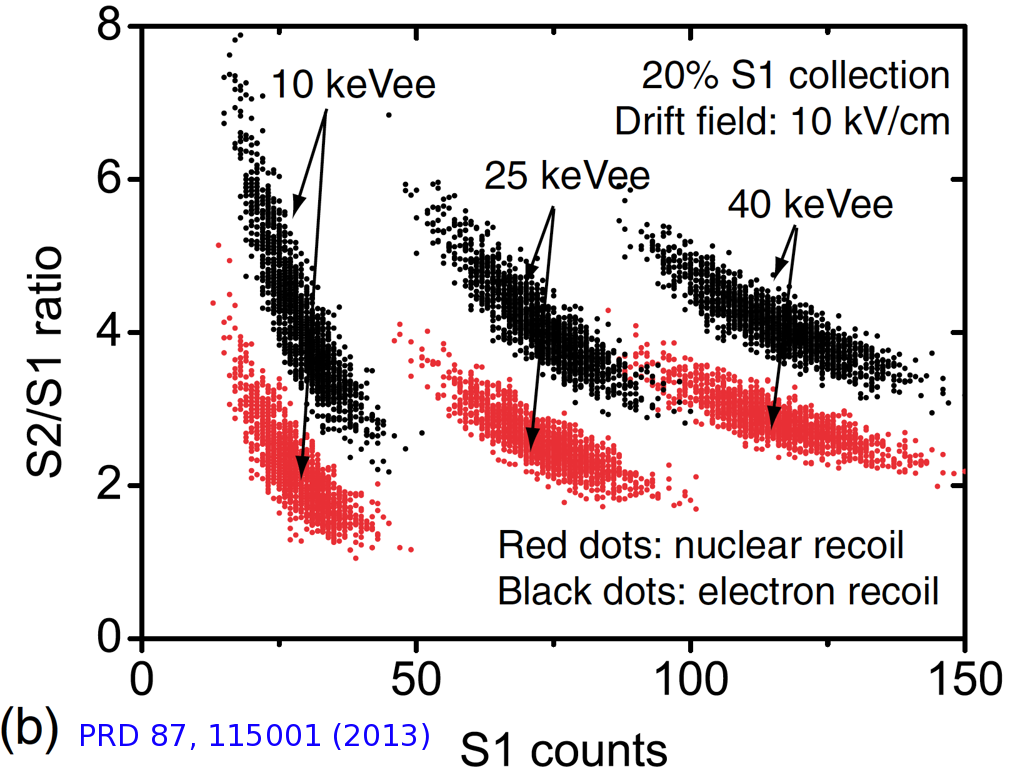}
		\caption{Fig.~\ref{fig_LHeS1S2.b}. ER and NR events are separated with a simulation of S1/S2 analysis. Fig.~\ref{fig_LHeS1S2.b} is copied from reference~\cite{ItoSeidel13}.}\label{fig_LHeS1S2.b}
	\end{subfigure}
	\caption{The calculated photons yield of LHe VS the electron and nuclear recoil energy of LHe (ER and NR). Fig.~\ref{fig_LHeS1S2.a} is the calculated absolute photons yield of ER and NR. Fig.~\ref{fig_LHeS1S2.b} is the relative yield: the yield of NR normalizes to the one of ER.}\label{fig_LHeS1S2}
\FloatBarrier
\end{figure}

The S2 signal resulted from dragging electrons away from the recombination process with an external electric field.  
Fig.~\ref{fig_LHeS1S2.a} indicates that ER and NR do not have the same fraction of elections collection for a certain electric filed, which could be implemented in the S1/S2 analysis for events discrimination. 

Fig.~\ref{fig_LHeS1S2.b} showed a good ER / NR discrimination in LHe for ionization energy down to 10 keV$_{\text{ee}}$ with the drift field of 10 kV/cm, and photo sensors can collect 20\% S1 scintillation~\cite{GuoMckinsey13}.

\captionsetup[subfigure]{labelformat=empty}

The drift field of LUX is 170 V/cm (LZ is 310 V/cm, and DarkSide-20k is 200 V/cm.). While for an LHe TPC, the drift field is required to be a factor of tens since  the electron collection is only 40\% even under 10 kV/cm, as shown in Fig.~\ref{fig_electrons.b}. Therefore, one of the critical R\&D programs for the ALETHEIA project is developing a safe and stable HV system up to 500 kV or higher.

\begin{figure}	
	\centering
	\begin{subfigure}[t]{3.0in}
		\centering
		\includegraphics[scale=0.22]{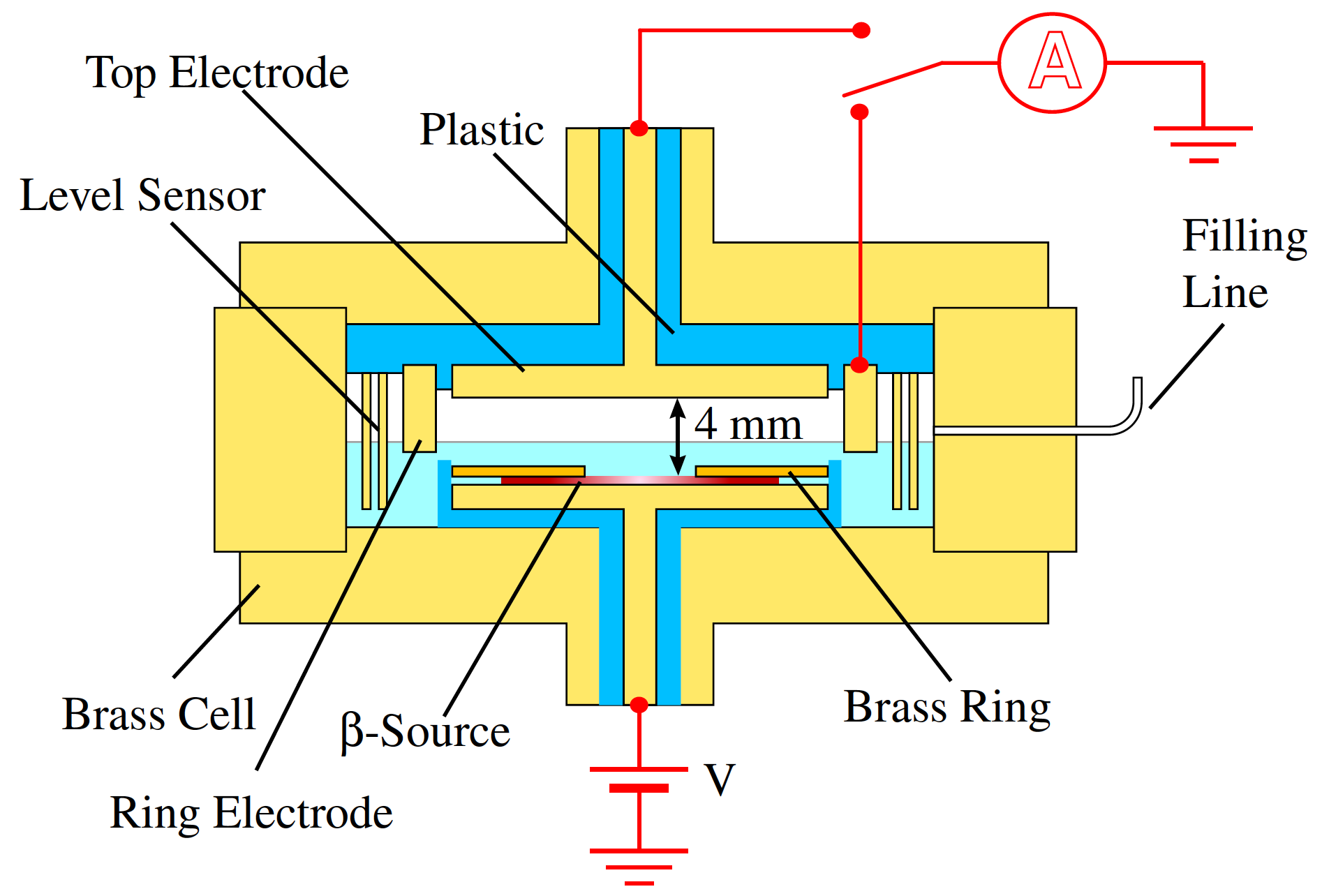}
		\caption{Fig.~\ref{fig_LHeS2Observation.a}. The schematic setup for the observation of electroluminescence in a liquid helium layer. Fig.~\ref{fig_LHeS2Observation.a} was copied from reference~\cite{SethumadhavanPhDThesis07}.}\label{fig_LHeS2Observation.a}	
		\end{subfigure}
	\quad
	\begin{subfigure}[t]{3.0in}
		\centering
		\includegraphics[scale=0.223, angle = 0]{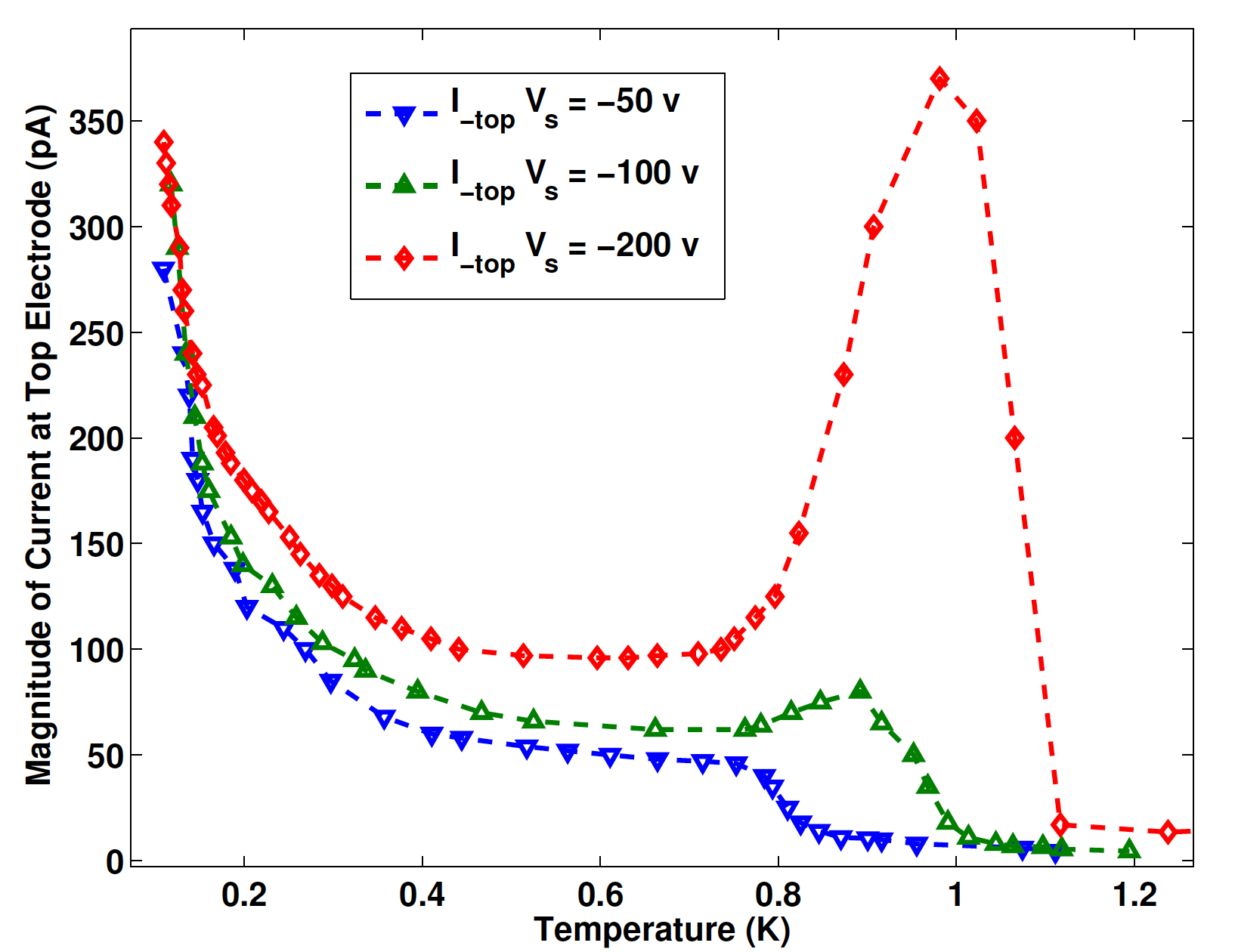}
		\caption{Fig.~\ref{fig_LHeS2Observation.b}. The intensity of electroluminescence changes with the applied field. Fig.~\ref{fig_LHeS2Observation.b} is copied from reference~\cite{SethumadhavanPhDThesis07}.}\label{fig_LHeS2Observation.b}
	\end{subfigure}
	\caption{Observation of electroluminescence in a small liquid helium cell under applied field.}\label{fig_LHeS2Observation}
\FloatBarrier
\end{figure}

Reference~\cite{SethumadhavanPhDThesis07} studied the production of electroluminescence in a dual-phase cell as shown schematically in Fig.~\ref{fig_LHeS2Observation.a}. This key part of the setup is a cylindrical cell with 0.4 cm in height and 2.5 cm in diameter. The top, bottom, and side of the cell connect to three different electrodes, respectively. The power source on the bottom provided the external voltage. The current on the tope and side electrodes can be measured by a current meter. For the test, the upper half of the cell is helium gas as the white area shown, the lower half in blue is liquid helium. A $^{63}$Ni beta source was put on the bottom of the cell. The electrons decayed from a $^{63}$N have the mean energy of 17 keV and the maximum energy of 66 keV. Electrons Decayed from the $^{63}$Ni will ionize liquid helium and produce ion-electron pairs. Under certain external fields, some electrons could drift upwards, extract from the liquid surface, and produce electroluminescence in the 2 mm helium gas. Fig.~\ref{fig_LHeS2Observation.a} shows under the voltage of $\sim$ 500 V/cm, significant electroluminescence current can generate. This test demonstrated a dual-phase TPC filled with helium could generate electroluminescence (or S2) signals, therefore, make the S1/S2 and S2O analysis of the ALETHEIA viable.

\captionsetup[subfigure]{labelformat=empty}

\subsection*{S2O for ALETHEIA}
The strategies of S2O analysis are not the same for DarkSide-50 and XENON-1T. For DarkSide-50, as mentioned in reference~\cite{DarkSide502018S2O}, to reach the lowest possible S2 signals, the fiducial volume can not be reconstructed with usual algorithm due to low photoelectron statistics (for S2), the fiducial region for the S2O analysis is in the $x-y$ plane by only accepting events where the largest S2 signal is recorded in one of the seven central top-array PMTs. For XENON-1T~\cite{XENON1TS2O19}, they used 30\% of Science Run (SR1) data as training data to determine events selections. Limits setting are computed using only the remaining 70\% data, which was not examined until the analysis was fixed. 

The critical feature to convince us that the S2O analysis makes sense is that the observed backgrounds are consistent with expected events for selected fiducial volume or datasets, as shown in Fig.~\ref{fig_3.a} and Fig.~\ref{fig_3.b}. 

The S2O analysis on the ALETHEIA can not be decided without data on hands. We will choose the most appropriate analysis once we have scientific data.

\subsection*{Critical calibrations with 30 g and 10 kg ALETHEIA prototype}
We briefly summarize the calibrations should be performed for the ALETHEIA TPC detector. \\
 A few versions of 30 g LHe cells would be designed to test: \\
Cal-I: ER and NR calibrations, \\
Cal-II: S2 signal optimizations, \\
Cal-III: SiPM testing at 4 K. \\

For the 10 kg prototype, the calibrations are:\\
Cal-IV: $\sim$ 500 kV or higher HV system work stably and safely with a LHe TPC ($\sim$ 10s- 100s of pA current in the LHe TPC.), \\
Cal-V: Given the slow drift speed of electrons in LHe ($\sim$ 2 m/s), a LHe TPC still be functional in a underground lab.\\



\subsection*{The progress of a 30 g ALETHEIA prototype detector}

\begin{figure}[!t]	 
	\centering
        \includegraphics[width=6.0in, angle = 0]{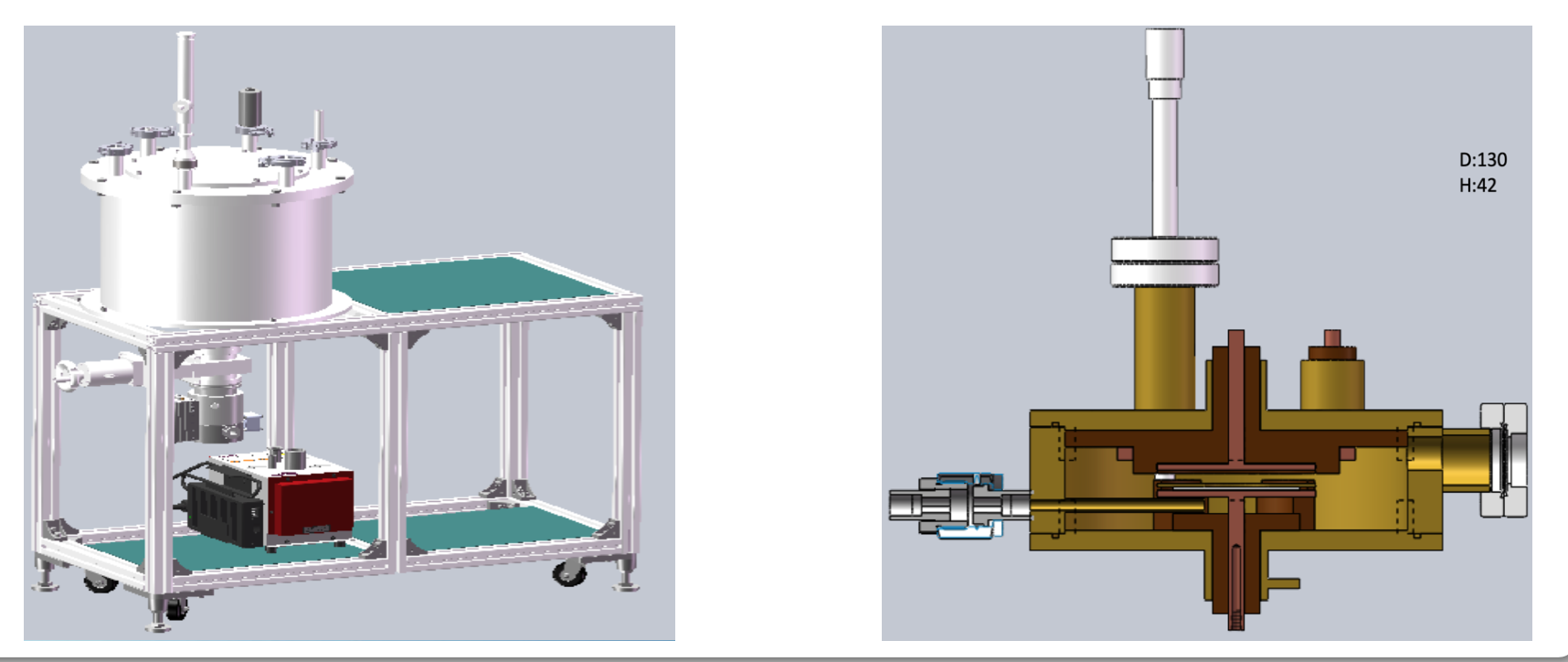}
	\caption{The mechanical drawing of our first 30 g LHe cell. The left plot shows the test bench. The right one is the internal structure of the whole 30 g LHe cell. The unit for ``D:130'' and ``H:42'' on the right plot is mm.}\label{30gCellV1ScehmaticDrawing}
\FloatBarrier
\end{figure}

We have designed our first 30 g LHe cell at CIAE (China Institute of Atomic Energy) in Beijing, China, as shown in Fig.~\ref{30gCellV1ScehmaticDrawing}. \\

\begin{figure}	
	\centering
	\begin{subfigure}[t]{3.1in}
		\centering
		\includegraphics[scale=0.28]{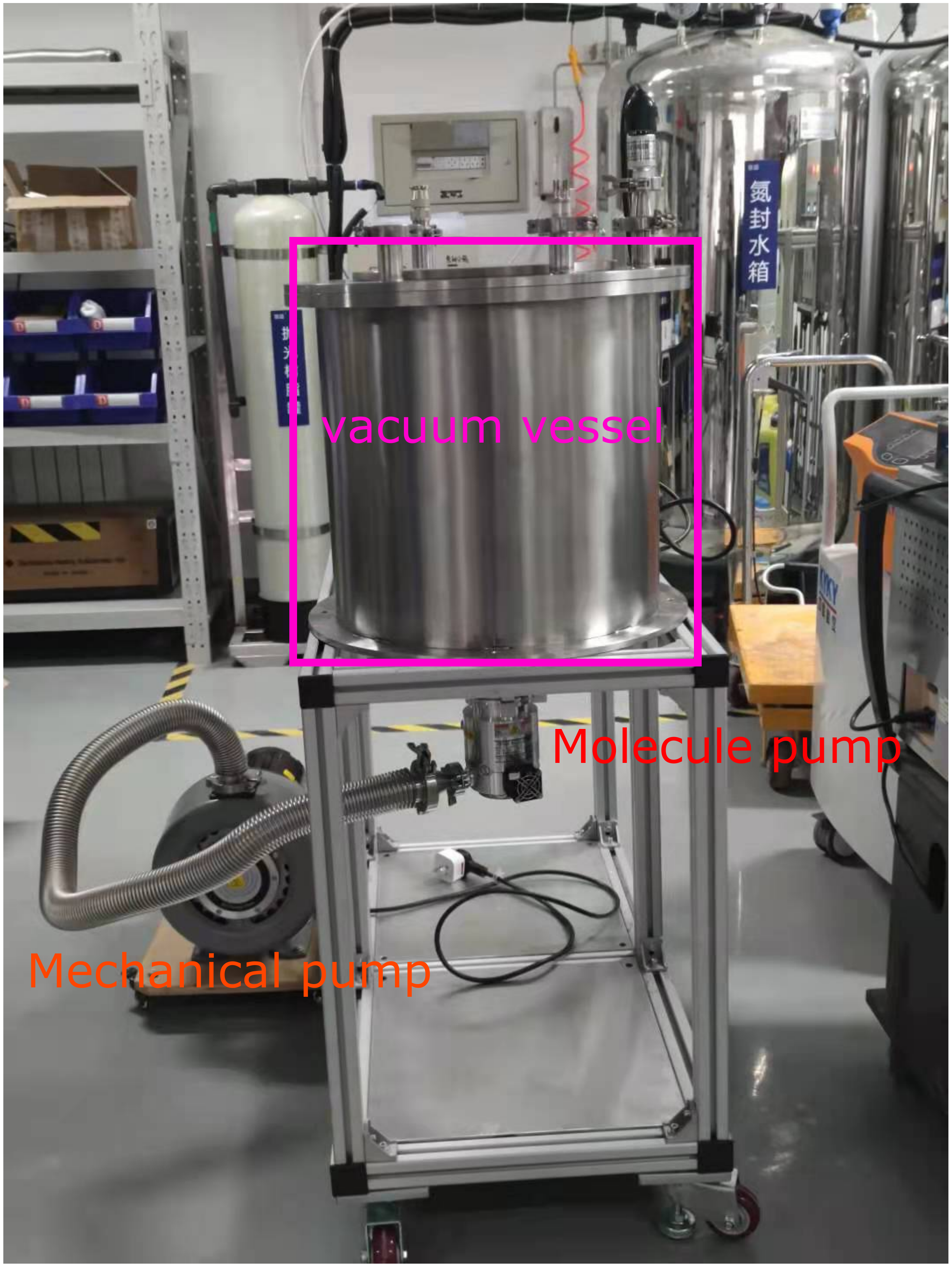}
		\caption{Fig.~\ref{fig_30gLHeCellAssembly.a}. The first 30 g LHe cell manufactured and assembled at CIAE.}\label{fig_30gLHeCellAssembly.a}	
		\end{subfigure}
	\quad
	\begin{subfigure}[t]{3.1in}
		\centering
		\includegraphics[scale=0.25, angle = 0]{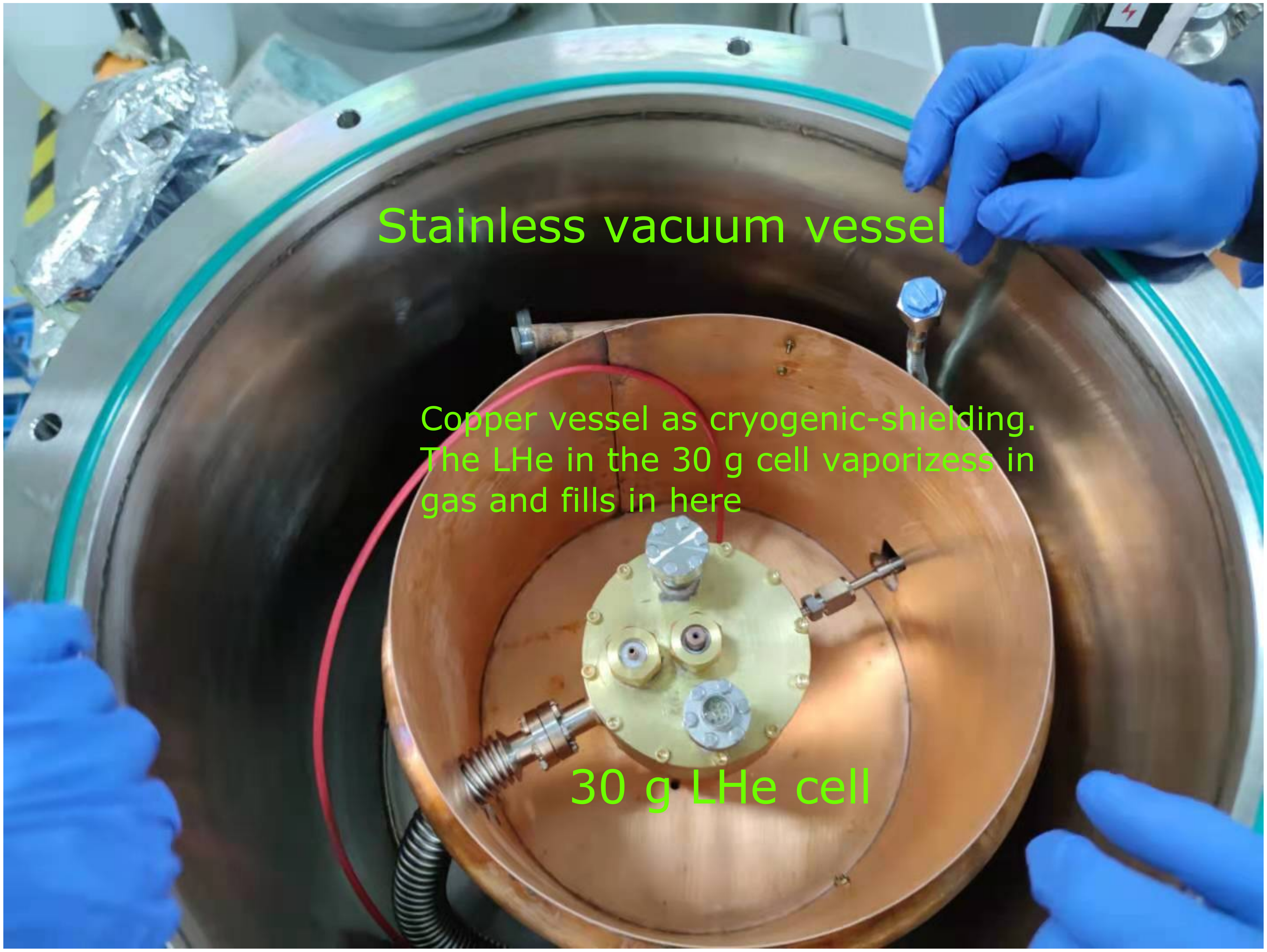}
		\caption{Fig.~\ref{fig_30gLHeCellAssembly.b}. The inside of the vacuum vessel as shown in Fig.~\ref{fig_30gLHeCellAssembly.a}.}\label{fig_30gLHeCellAssembly.b}
	\end{subfigure}
	\quad
		\begin{subfigure}[t]{3.1in}
		\centering
		\includegraphics[scale=0.15]{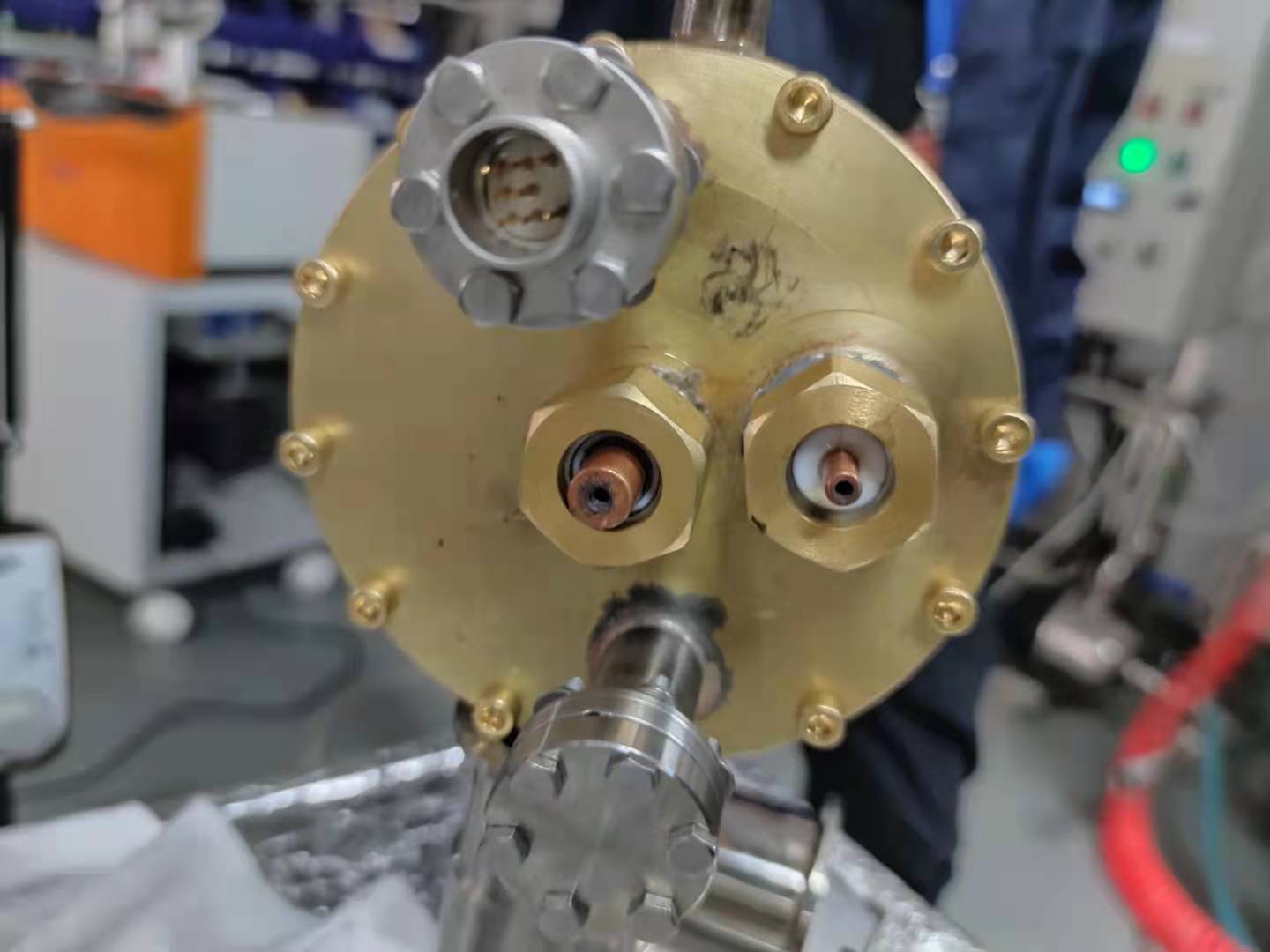}
		\caption{Fig.~\ref{fig_30gLHeCellAssembly.c}. The 30 g LHe cell as shown in the center of Fig.~\ref{fig_30gLHeCellAssembly.b}.}\label{fig_30gLHeCellAssembly.c}	
		\end{subfigure}
	\quad
	\begin{subfigure}[t]{3.1in}
		\centering
		\includegraphics[scale=1.05, angle = 0]{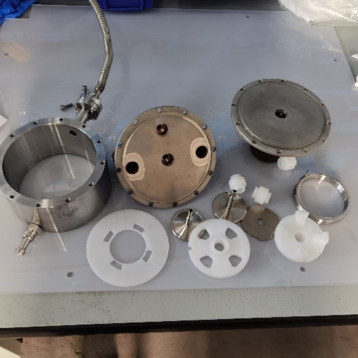}
		\caption{Fig.~\ref{fig_30gLHeCellAssembly.d}. The parts of the 30 g LHe cell as shown in Fig.~\ref{fig_30gLHeCellAssembly.c}.}\label{fig_30gLHeCellAssembly.d}
	\end{subfigure}

	\caption{The first 30 g LHe cell was built at CIAE (home made) in Beijing, China.}\label{fig_30gLHeCellAssembly}
\FloatBarrier
\end{figure}

Fig.~\ref{fig_30gLHeCellAssembly} showed the first version of our 30 g LHe cell assembled at CIAE. The primary purpose of this detector is to gain experience in building an apparatus capable of working at LHe temperature ($\sim$ 4.5 K), make sure external DC High Voltage (HV) could be applied to the detector, and the dark current of the detector under HV should be small enough.

\captionsetup[subfigure]{labelformat=empty}

\begin{figure}	
	\centering
	\begin{subfigure}[t]{3.1in}
		\centering
		\includegraphics[scale=0.4]{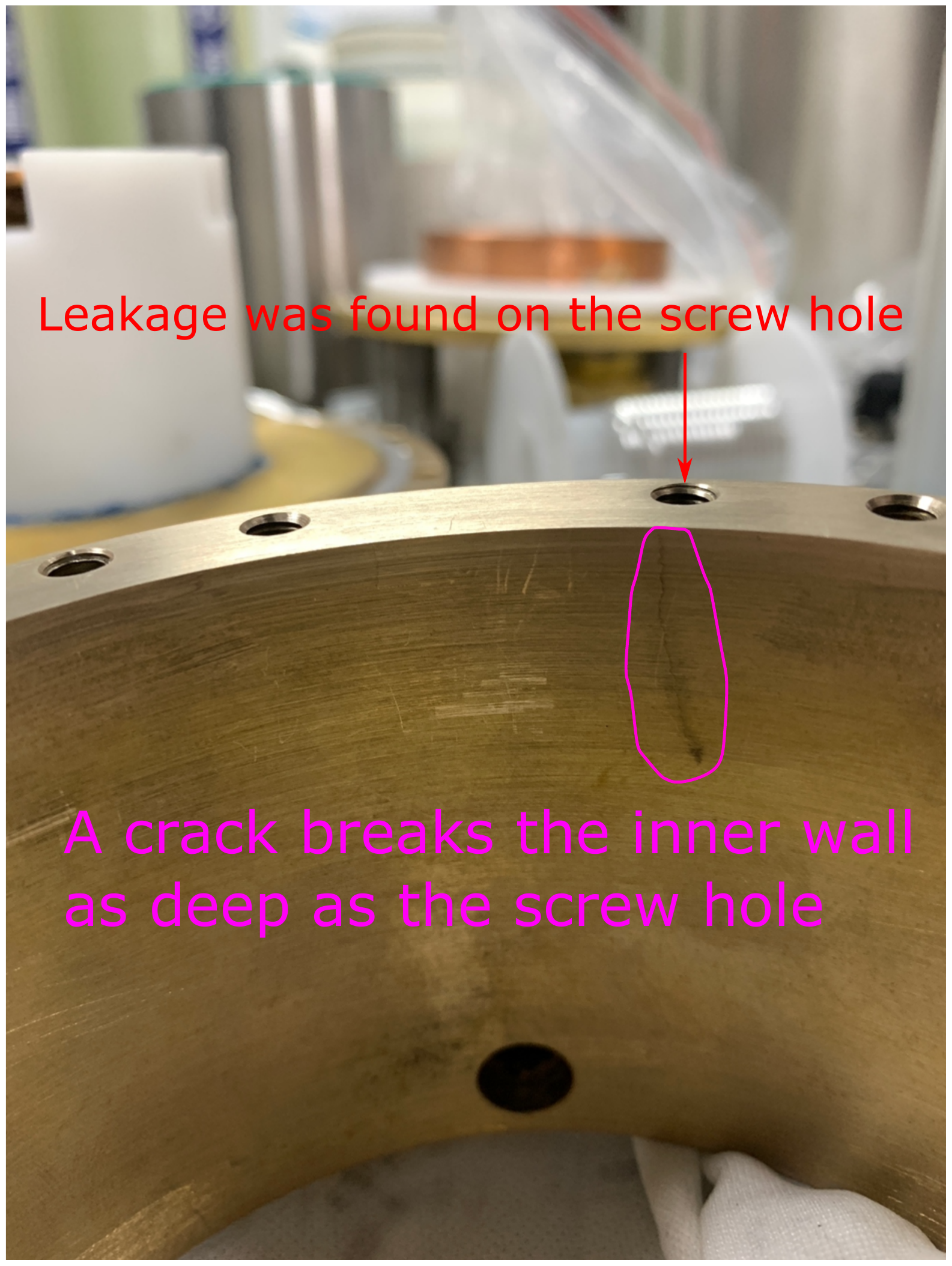}
		\caption{Fig.~\ref{fig_30gLHeVacuumLeakage.a}. An invisible crack on the inner wall of the 30 g LHe detector breaks the wall paving the way of vacuum leakage to the screw hole.}\label{fig_30gLHeVacuumLeakage.a}	
		\end{subfigure}
	\quad
	\begin{subfigure}[t]{3.1in}
		\centering
		\includegraphics[scale=1.25, angle = 0]{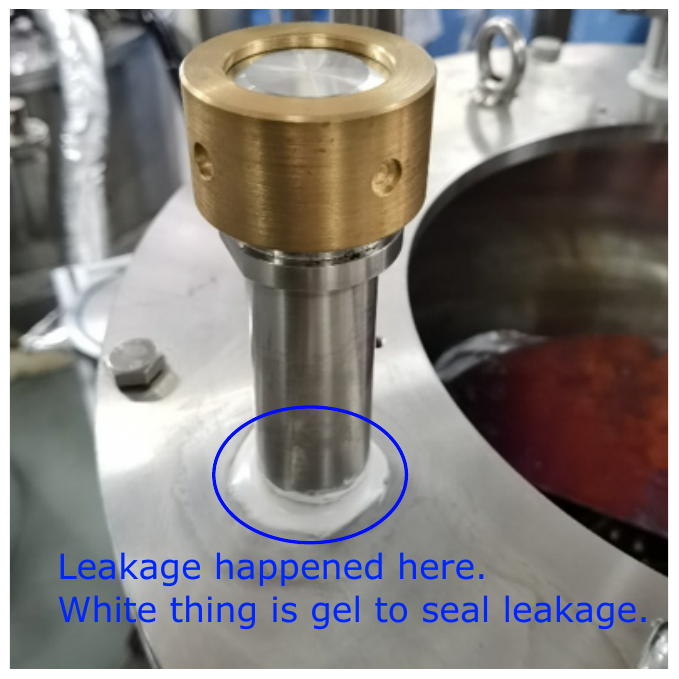}
		\caption{Fig.~\ref{fig_30gLHeVacuumLeakage.b}. The welding place of the LHe inlet on the vacuum vessel had a leakage. Vacuum gel was used to seal the leakage successfully.}\label{fig_30gLHeVacuumLeakage.b}
	\end{subfigure}
	\caption{Two examples of vacuum leakage observed on the 30 g LHe cell we built at CIAE.}\label{fig_30gLHeVacuumLeakage}
\FloatBarrier
\end{figure}

To cool the 30 g detector (as shown in Fig.~\ref{fig_30gLHeCellAssembly.c}) down to 4.5 K, one must first check the vacuum of the whole detector system is good enough or not; if not, the vacuum-leaking places would lead to cryogenic leakage either. Consequently, the detector can not reach the LHe temperature. Even if the design of the detector is perfect, vacuum leakage can still happen anywhere due to mechanical failures; some of them are difficult to track. We have found more than a handful of leaking places in building our first detector. We only show two of them here as examples. As indicated in Fig.~\ref{fig_30gLHeVacuumLeakage.a}, we detected vacuum leakage at the screw hole place but couldn't figure out how the leakage could happen here since everything looked pretty good until we found a tiny crack (when we performed vacuum tests, a cover sealed on it with screws as shown in Fig~\ref{fig_30gLHeCellAssembly.c}). It turns out the almost invisible tiny crack as shown in Fig.~\ref{fig_30gLHeVacuumLeakage.a} breaks the inner wall of the detector all the way down to the screw hole. The whole leaking routine is: the inside of the cell $\rightarrow$ the tiny crack $\rightarrow$ the wall $\rightarrow$ the screw hole. To solve the problem, we made a new 30 g detector made of stainless 304 (instead of copper) thanks to its better mechanical performance. Fig.~\ref{fig_30gLHeVacuumLeakage.b} shows another leakage that happened on the welding place of the LHe inlet on the vacuum vessel due to a welding failure. As shown in the plot, vacuum gel was used to seal the leakage. With the gel, no vacuum leakage happens anymore here.

\captionsetup[subfigure]{labelformat=empty}

\begin{figure}	
	\centering
	\begin{subfigure}[t]{3.1in}
		\centering
		\includegraphics[scale=0.2]{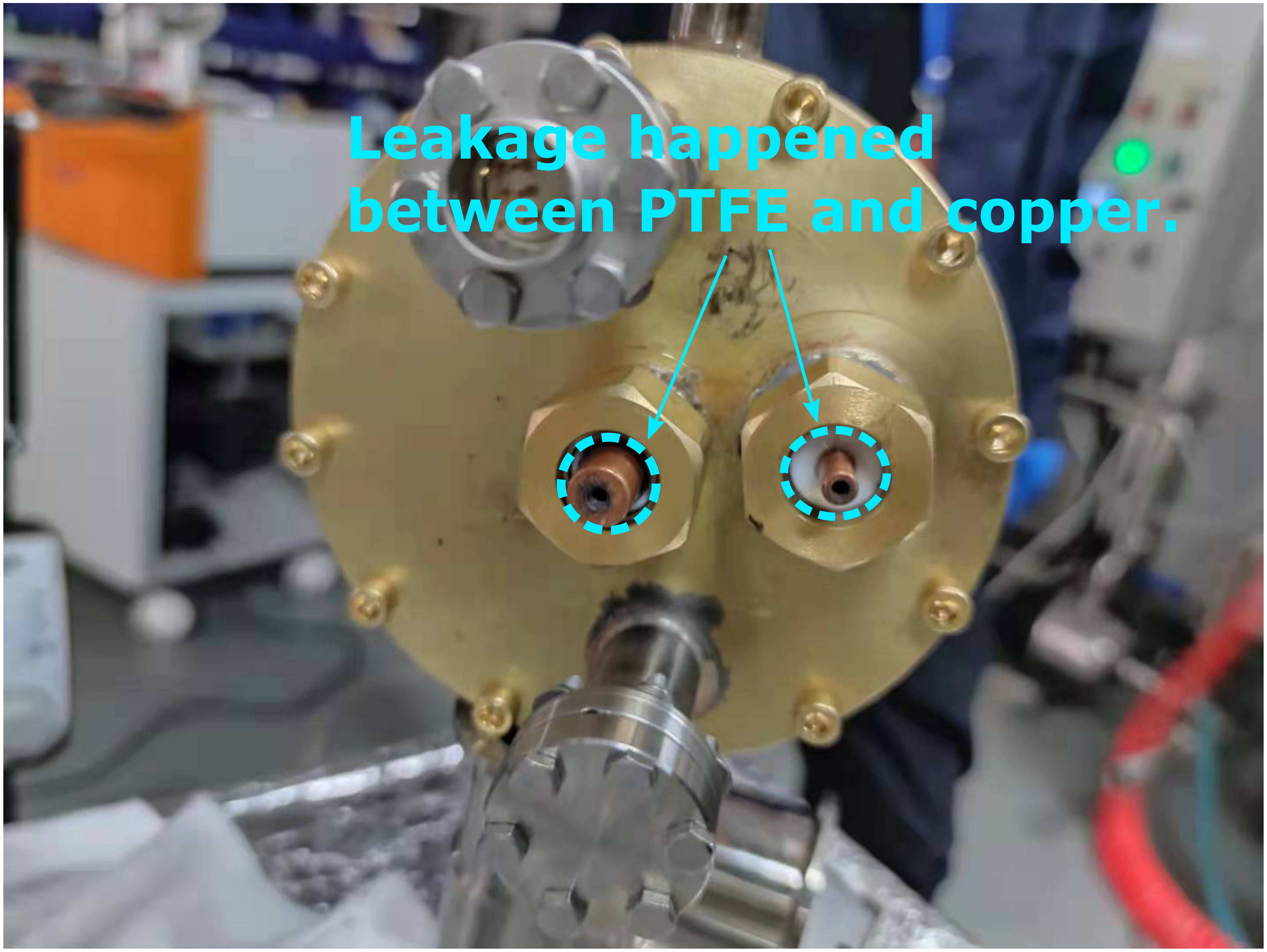}
		\caption{Fig.~\ref{fig_30gLHeCryogenicLeakage.a}. Leakage was detected at the cyan circular places when the detector was cooled with liquid nitrogen. Refer to texts for explanations.}\label{fig_30gLHeCryogenicLeakage.a}	
		\end{subfigure}
	\quad
	\begin{subfigure}[t]{3.1in}
		\centering
		\includegraphics[scale=0.054, angle = 0]{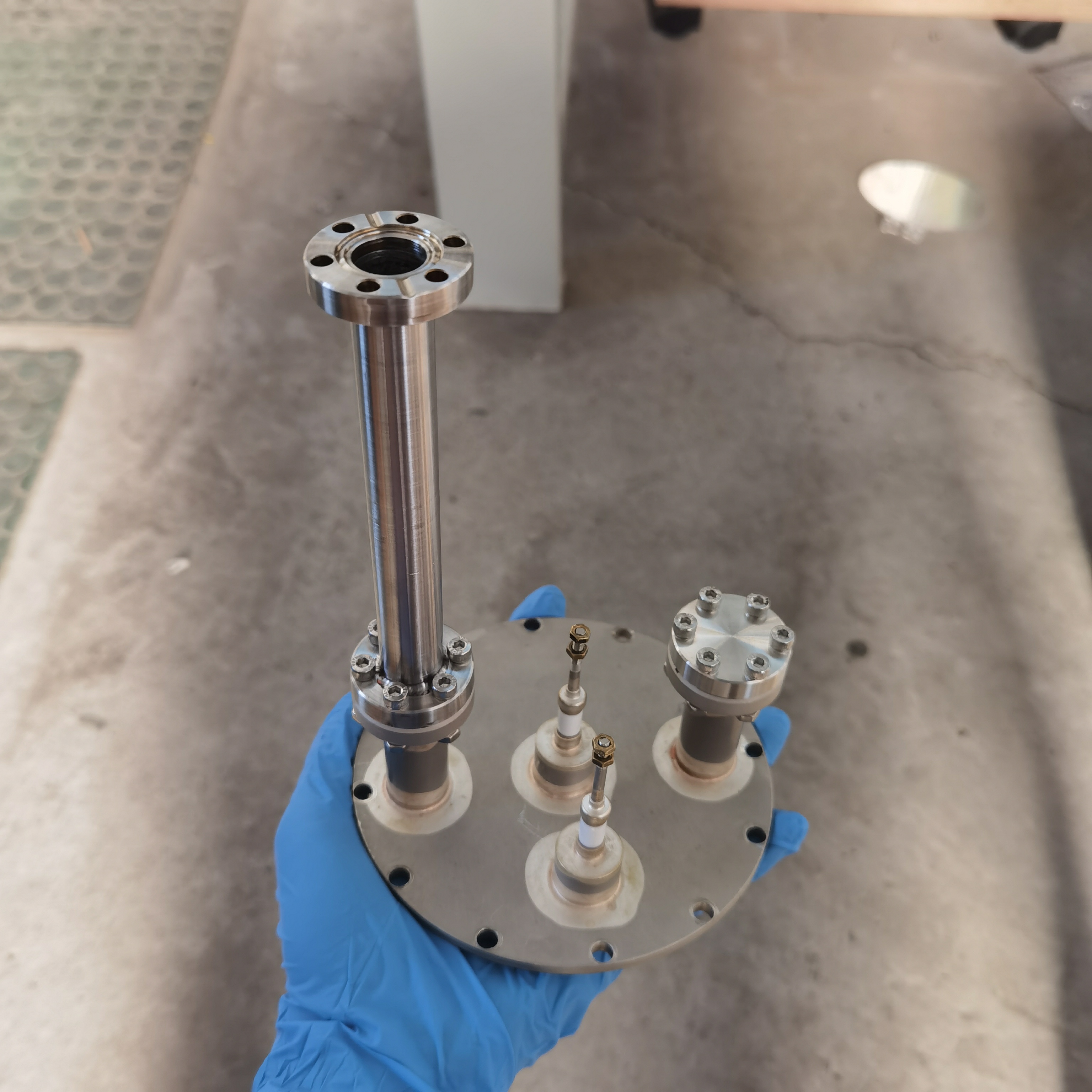}
		\caption{Fig.~\ref{fig_30gLHeCryogenicLeakage.b}.  The ceramic electrode we ordered to cope with the cryogenic leakage shown in Fig~.\ref{fig_30gLHeCryogenicLeakage.a}. With this board, our detector was able to cool to LHe temperature successfully.}\label{fig_30gLHeCryogenicLeakage.b}
	\end{subfigure}
	\caption{One example of cryogenic leakage observed on the 30 g LHe cell we built at CIAE. A ceramic electrode works well down to LHe temperature.}\label{fig_30gLHeCryogenicLeakage}
\FloatBarrier
\end{figure}

However, even if there is no vacuum leakage anymore in the detector system at room temperature, it is still impossible for the detector to achieve the LHe temperature if cryogenic leakage exists. Here, cryogenic leakage represents the leakage that only happened in the process of cooling the detector. We had a few lessons of this kind. We only show two of them here as examples. The first cryogenic leakage happened when we cooled the detector with Liquid Nitrogen (LN). We located the leakage at two electrodes (high voltage anode and cathode) and realized the reason for the leakage is due to different Coefficients of Linear Thermal Expansion (CLTE) of the materials we implemented to build the electrodes. Both electrodes were made in the same structure though their dimensions are slightly different. A copper pin is in the center, which is surrounded by an isolating PTFE cylinder; the PTFE layer then connects to the detector's inner wall, as shown in Fig.~\ref{fig_30gLHeCellAssembly.c}. The electrodes worked well at room temperature but leaked as long as we cooled them with LN. The CLTE of copper at LN temperature (77 K) is 0.302 (compared to room temperature, 293 K.), while PTFE is 1.984. Thus, there exists a difference of more than six-fold. So, when the detector was cooled to LN temperature, the PTFE isolator shrunk six times more than the copper wall.
Consequently, a tiny gap exists between the PTFE and the inner wall of the detector, as shown in Fig~.\ref{fig_30gLHeCryogenicLeakage.a}. Which eventually leads to leakage. To solve the problem and speed up the progress, we outsourced a ceramic electrode, as shown in Fig~.\ref{fig_30gLHeCryogenicLeakage.b}. With the electrode (and the new stainless 30 g detector), we finally were capable of cooling the detector down to LN temperature. 

\captionsetup[subfigure]{labelformat=empty}

\begin{figure}	
	\centering
	\begin{subfigure}[t]{3.1in}
		\centering
		\includegraphics[scale=0.2]{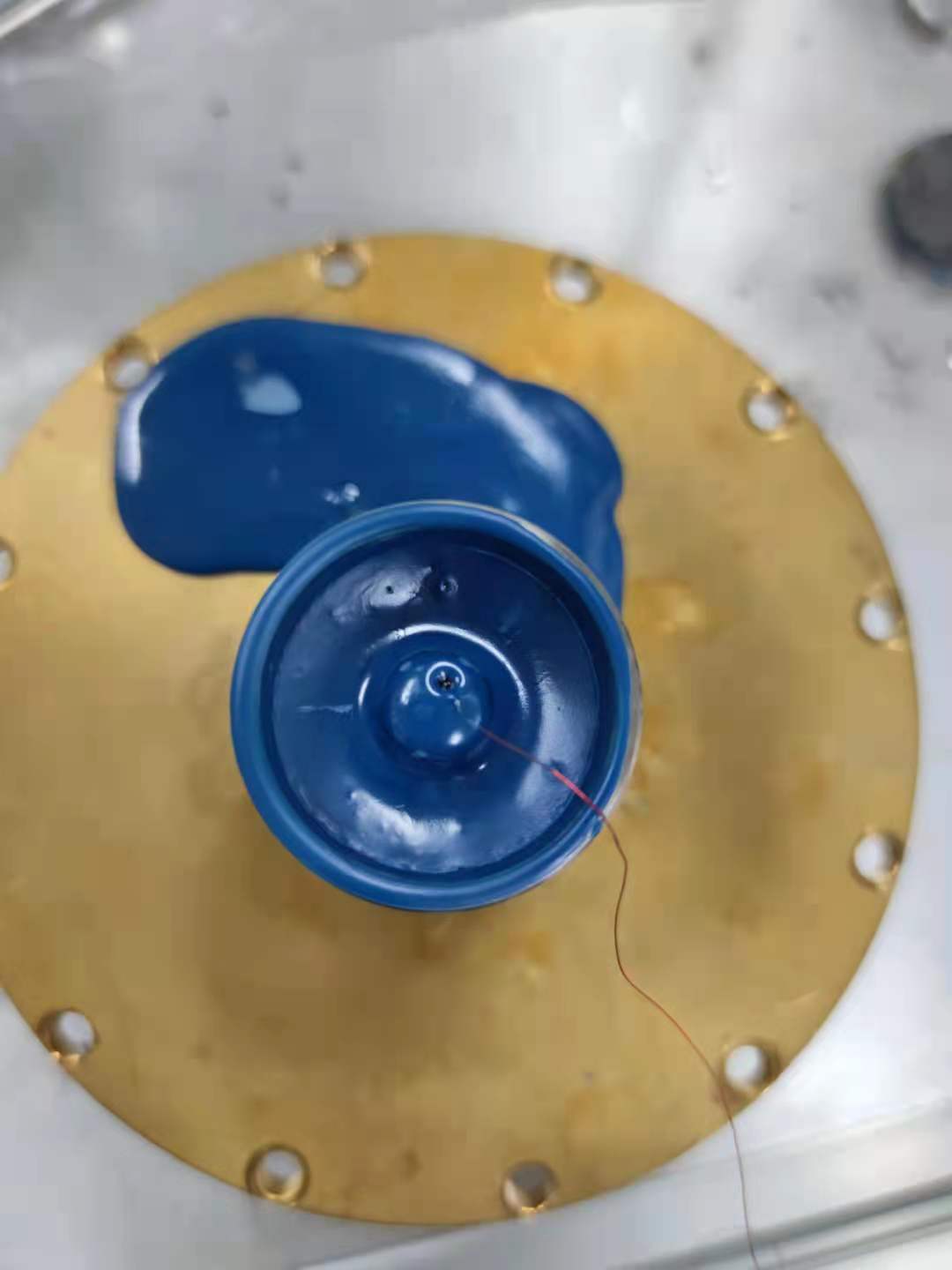}
		\caption{Fig.~\ref{fig_30gLHeCryogenicLeakage2.a}. The electrode encapsulated  with ``Loctite Staycast 8250FT Blue + CAT 11'' does not have leakage at LN temperature.}\label{fig_30gLHeCryogenicLeakage2.a}	
		\end{subfigure}
	\quad
	\begin{subfigure}[t]{3.1in}
		\centering
		\includegraphics[scale=0.15, angle = 0]{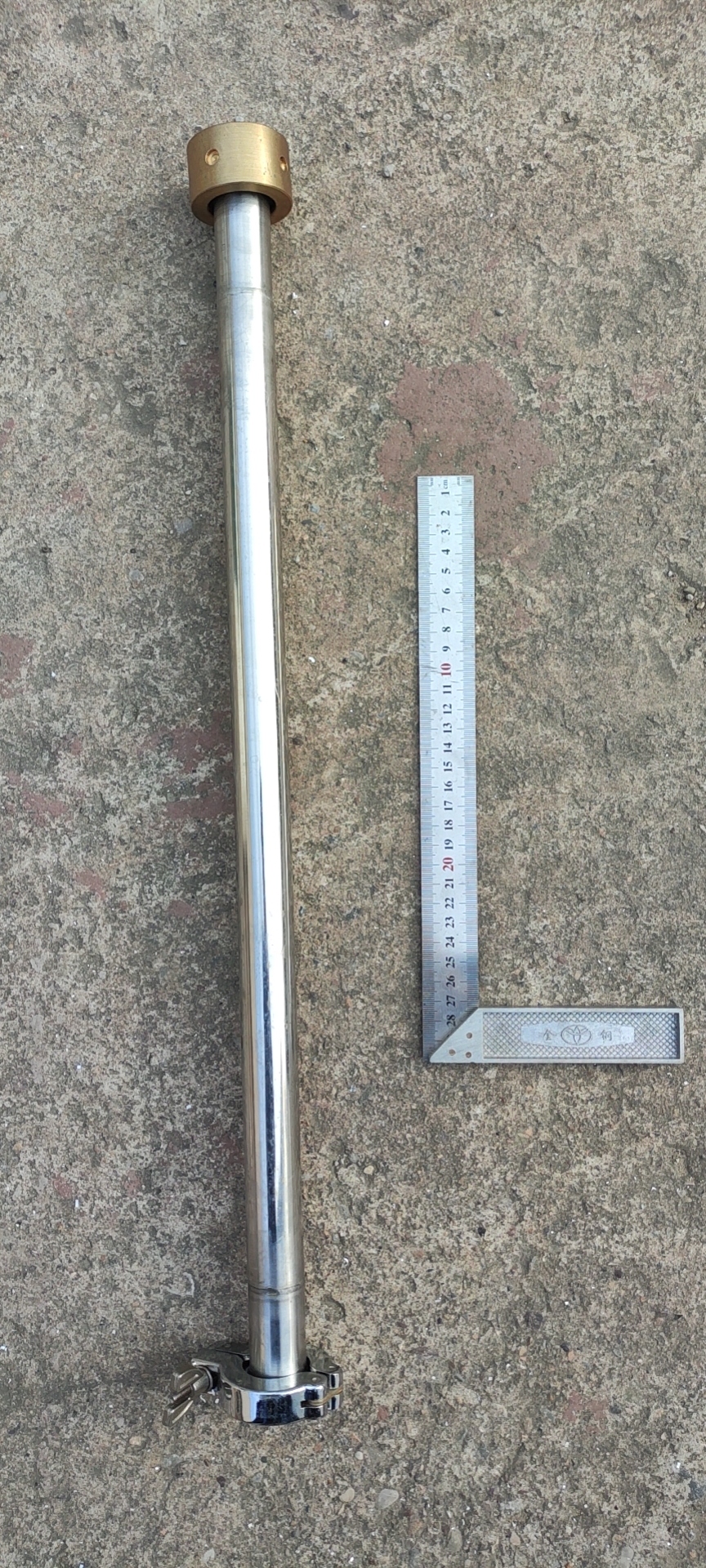}
		\caption{Fig.~\ref{fig_30gLHeCryogenicLeakage2.b}. The LHe transport tube doesn't have a vacuum layer inside. With it, the detector can only cool to $\sim$ 10.0 K.}\label{fig_30gLHeCryogenicLeakage2.b}
	\end{subfigure}
	\caption{Another example of cryogenic leakage.}\label{fig_30gLHeCryogenicLeakage2}
\FloatBarrier
\end{figure}

In parallel, we tried other possible solutions to cope with this kind of cryogenic issue. One of the successful tries is that we encapsulated the whole electrode with Loctite Staycast 8250FT Blue + CAT 11~\cite{8250FTBlueTDS}, as shown in Fig.~\ref{fig_30gLHeCryogenicLeakage2.a}. With this electrode, our 30 g detector is capable of working at LN temperature~\footnote{We thank Prof. George Seidel at Brown and Dr. Takeyasu Ito at LANL for sharing their experiences on Loctite Staycast 8250FT Blue + CAT 11.}. We will continue to design and home-make electrodes to satisfy our requirements: capable of working at LHe temperature and conveying 100s of kV voltage or higher, which are necessary for 100s kg or ton-scale of LHe TPCs.

Another relatively smaller cryogenic leakage happened on the tube where LHe can flow from the liquefying plant to our detector. The initial version of the tube does not have a vacuum layer inside, as shown in Fig.~\ref{fig_30gLHeCryogenicLeakage2.b}. With such a tube equipped, the lowest temperature the detector can reach is $\sim$ 10.0 K. We later replaced it with a tube containing a 2 mm vacuum layer inside. Until then, the detector can be cooled to a temperature lower than 10 K.

\captionsetup[subfigure]{labelformat=empty}

\begin{figure}[!t]	 
	\centering
        \includegraphics[width=6.0in, angle = 0]{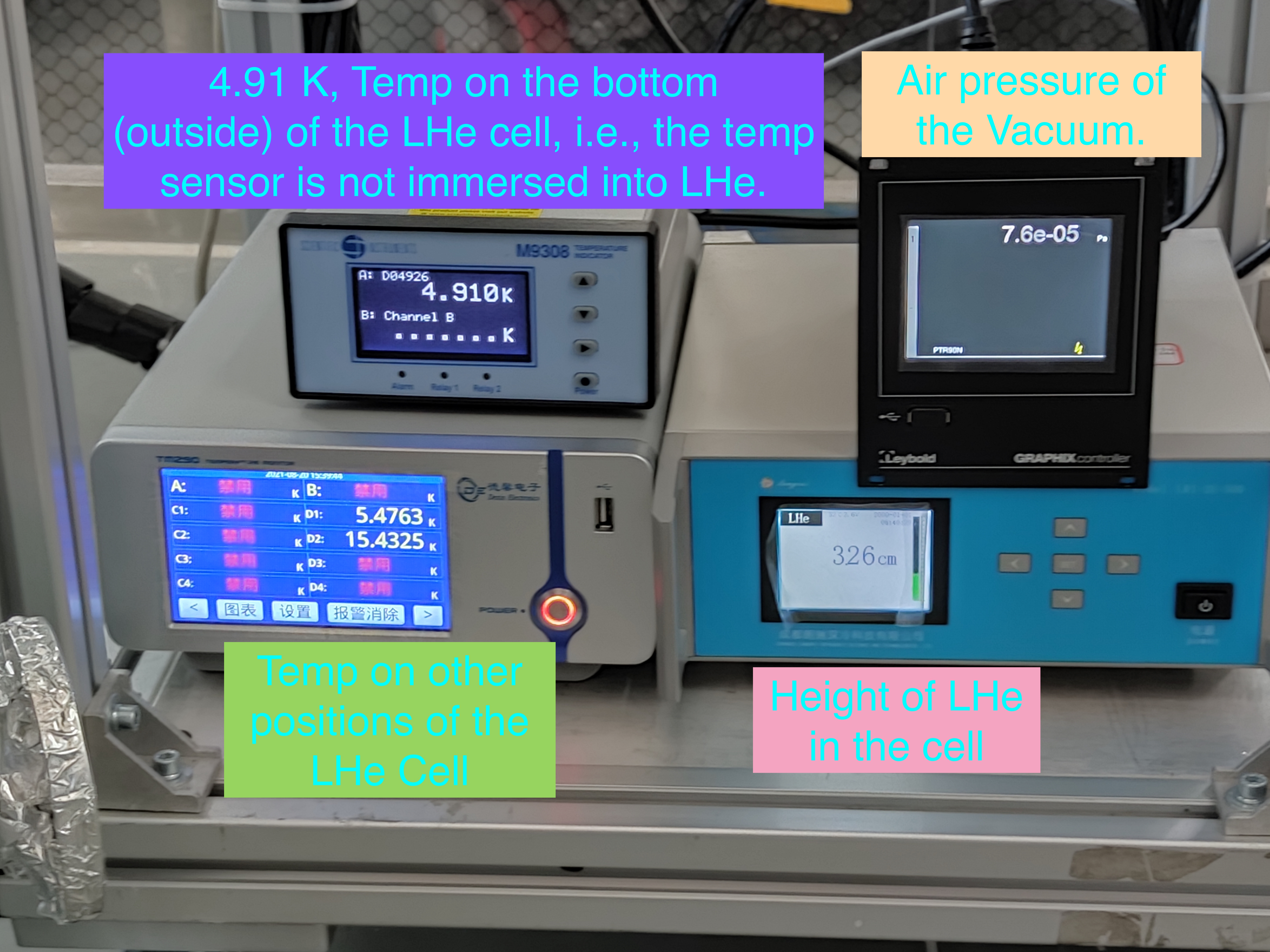}
	\caption{The screen on the upper left shows the temperature on the bottom side of the 30 g detector (The temperature sensor did not immerse into the LHe.), so it's slightly higher than 4.5 K; however, the screen on the lower right clearly shows the height of LHe in the cell is 3.26 cm. The upper right screen is the vacuum inside of the stainless vacuum vessel. The lower left screen shows the temperature on the top side of the 30 g cell (5.4763 K) and the helium gas in the copper vessel (15.4325 K), respectively. Please refer to Fig.~\ref{fig_30gLHeCellAssembly.d}.}\label{LHeTemperatureAchieved}
\FloatBarrier
\end{figure}

After many efforts have been made, we ultimately cooled our detector down to LHe temperature in summer 2021, as shown in Fig.~\ref{LHeTemperatureAchieved}.

\begin{figure}	
	\centering
	\begin{subfigure}[t]{3.1in}
		\centering
		\includegraphics[scale=0.4]{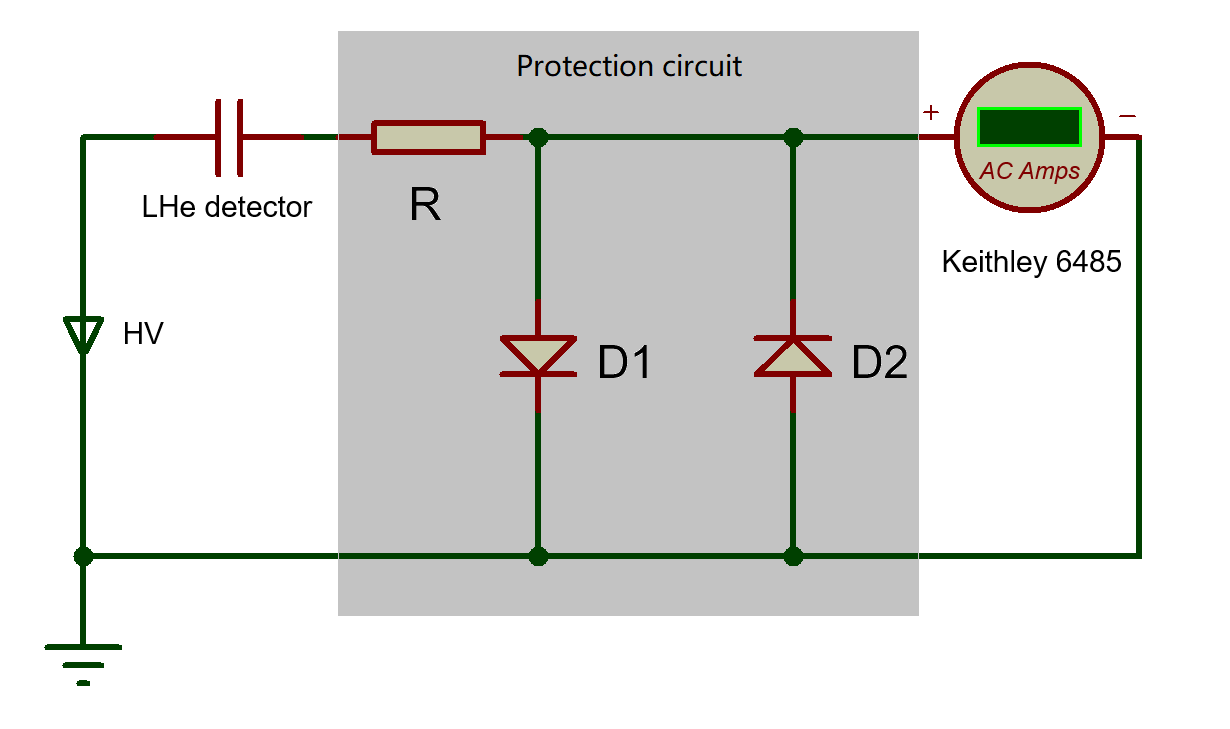}
		\caption{Fig.~\ref{fig_30gLHeDarkCurrentTest.a}. The schematic drawing of dark current tests on the 30 g LHe detector.}\label{fig_30gLHeDarkCurrentTest.a}	
		\end{subfigure}
	\quad
	\begin{subfigure}[t]{3.1in}
		\centering
		\includegraphics[scale=0.5, angle = 0]{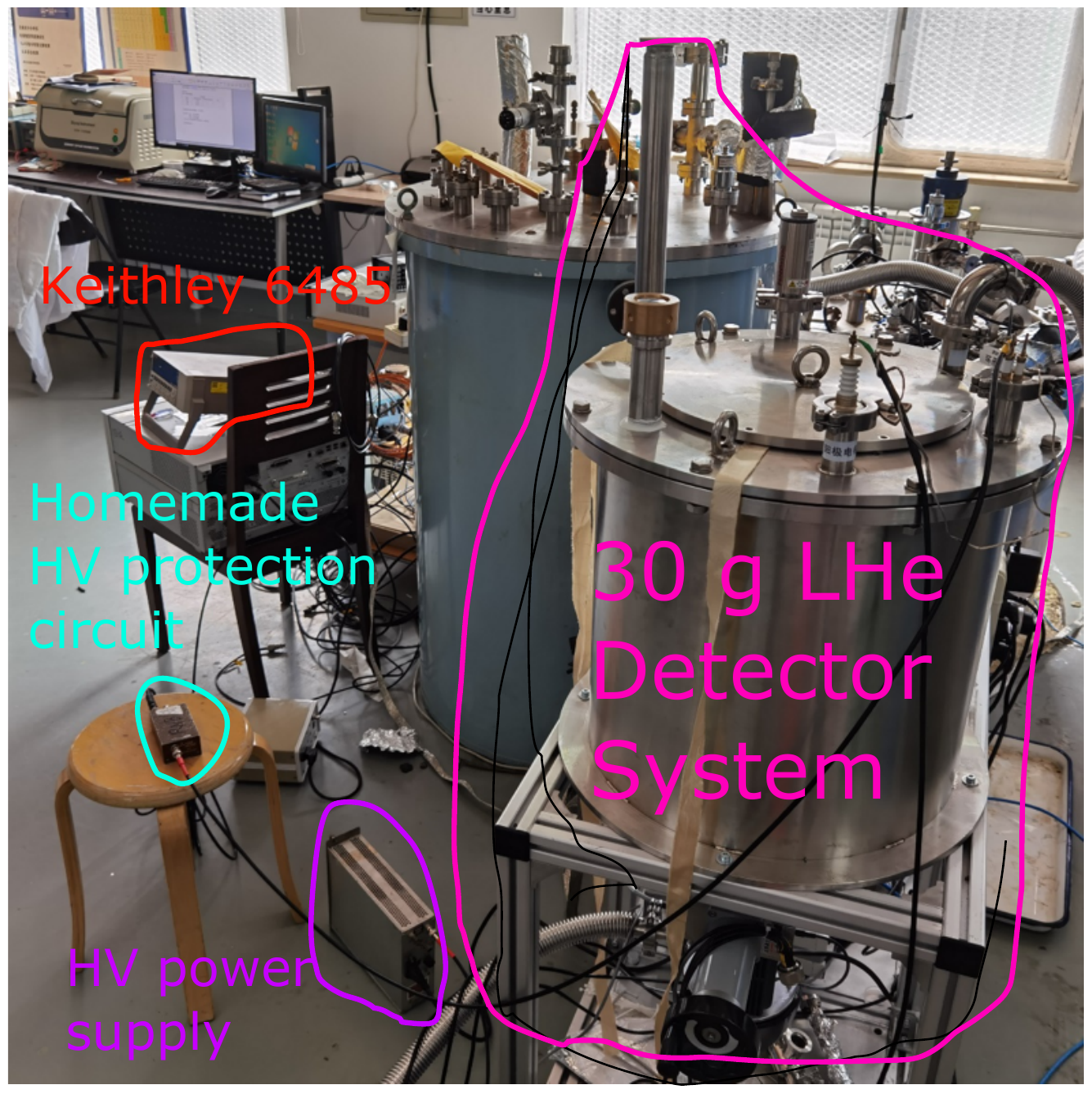}
		\caption{Fig.~\ref{fig_30gLHeDarkCurrentTest.b}. The experimental setup of dark current tests on the 30 g LHe detector.}\label{fig_30gLHeDarkCurrentTest.b}
	\end{subfigure}
	\caption{Another example of cryogenic leakage.}\label{fig_30gLHeDarkCurrentTest}
\FloatBarrier
\end{figure}

After the detector is capable of working at LHe temperature, our original plan was to characterize the detector by measuring the current produced by the ionized particles of a Ni-63 beta source, which would be immersed in the LHe, as indicated in references~\cite{Seidel14}. The electrons emitted from the Ni-63 source have an average energy of 17.5 keV. Electrons ionize LHe and generate electron-ion pairs. Under external HV, ionized electrons and ions move towards electrodes, generate current along with their movement, which can be registered in an electrometer. However, in September 2021, the liquefying system we relied on for the LHe supplement was out of work due to the damage of a helium gas recycling bag. So, we implement the same setup to test the dark current of the LHe system when filled with vacuum, nitrogen gas, and liquid nitrogen. Fig.~\ref{fig_30gLHeDarkCurrentTest.a} shows the schematic drawing of the current test setup. Fig.~\ref{fig_30gLHeDarkCurrentTest.b} is the experimental setup. The HV protection circuit in Fig.~\ref{fig_30gLHeDarkCurrentTest.b} is homemade at CIAE. The detailed circuit is show in Fig.~\ref{fig_30gLHeDarkCurrentTest.a}. The resistor is 1 G$\Omega$, and the diodes are both 1N3595. The resistor and diodes are all low-noise elements. The electrometer is Keithley 6485. For the HV power supply, we implemented ORTEC 556 and CAEN NDT1470. 

\captionsetup[subfigure]{labelformat=empty}

Before testing the detector's dark current, we cleaned up all of the parts of the 30 g LHe cell (as shown in Fig.~\ref{fig_30gLHeCellAssembly.d}) with the following procedure. \\
$\bullet$ Clean all parts 30 minutes with purified water and dishwashing liquid in a supersonic wave machine. \\
$\bullet$ Clean 30 minutes with purified water only in the supersonic wave machine.\\
$\bullet$ Flush with a high-pressure water gun (purified water).\\
$\bullet$ Wipe all parts' surfaces with alcohol.\\
$\bullet$ Blow with nitrogen gas.\\
$\bullet$ Dry with a vacuum drying machine.\\
$\bullet$ Clean electrical parts such as electrodes with a plasma gun.\\

\begin{figure}	
	\centering
	\begin{subfigure}[t]{3.1in}
		\centering
		\includegraphics[scale=0.35]{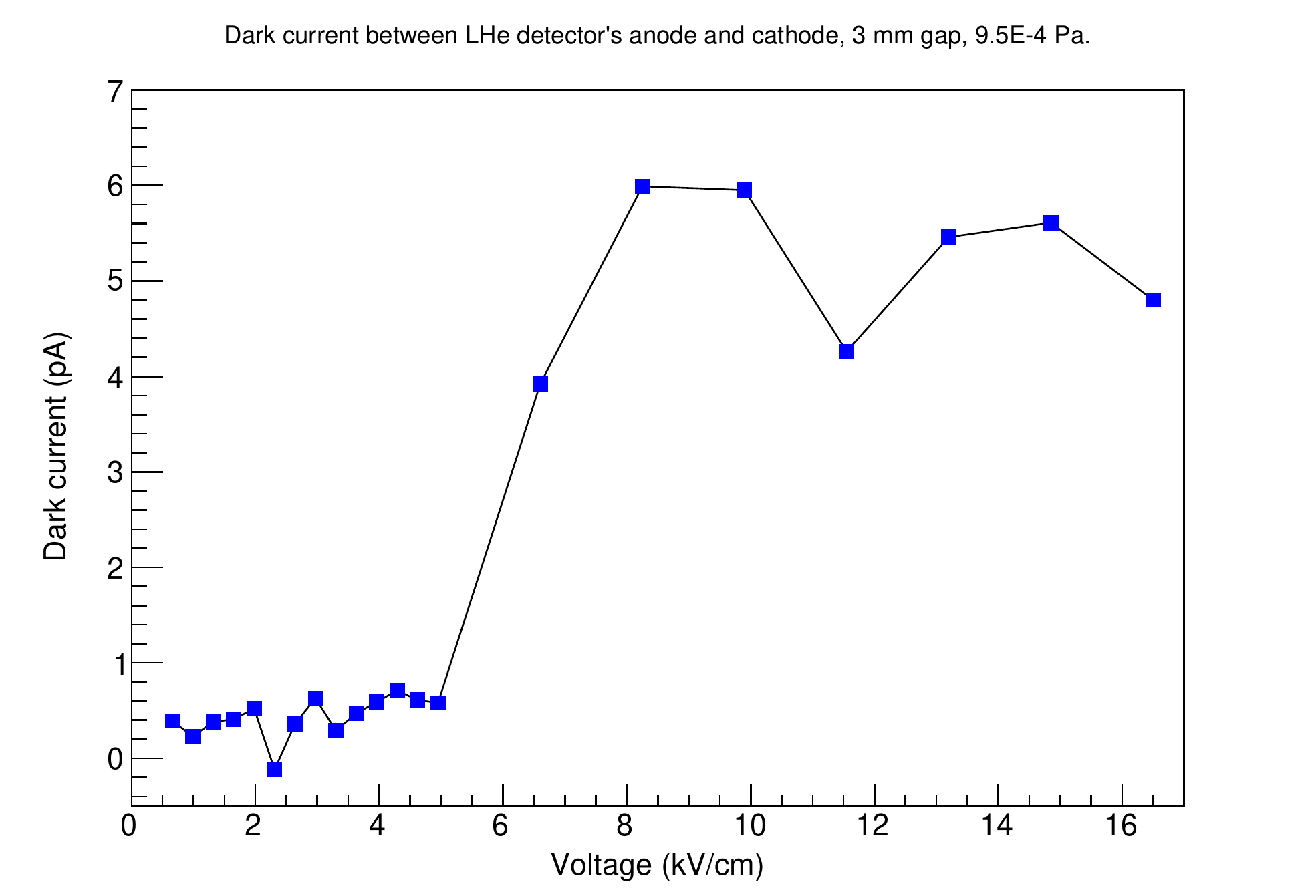}
		\caption{Fig.~\ref{fig_30gLHeDarkCurrentResults.a}. The dark current of the 30 g LHe cell, measured under vacuum.}\label{fig_30gLHeDarkCurrentResults.a}	
		\end{subfigure}
	\quad
	\begin{subfigure}[t]{3.1in}
		\centering
		\includegraphics[scale=0.35, angle = 0]{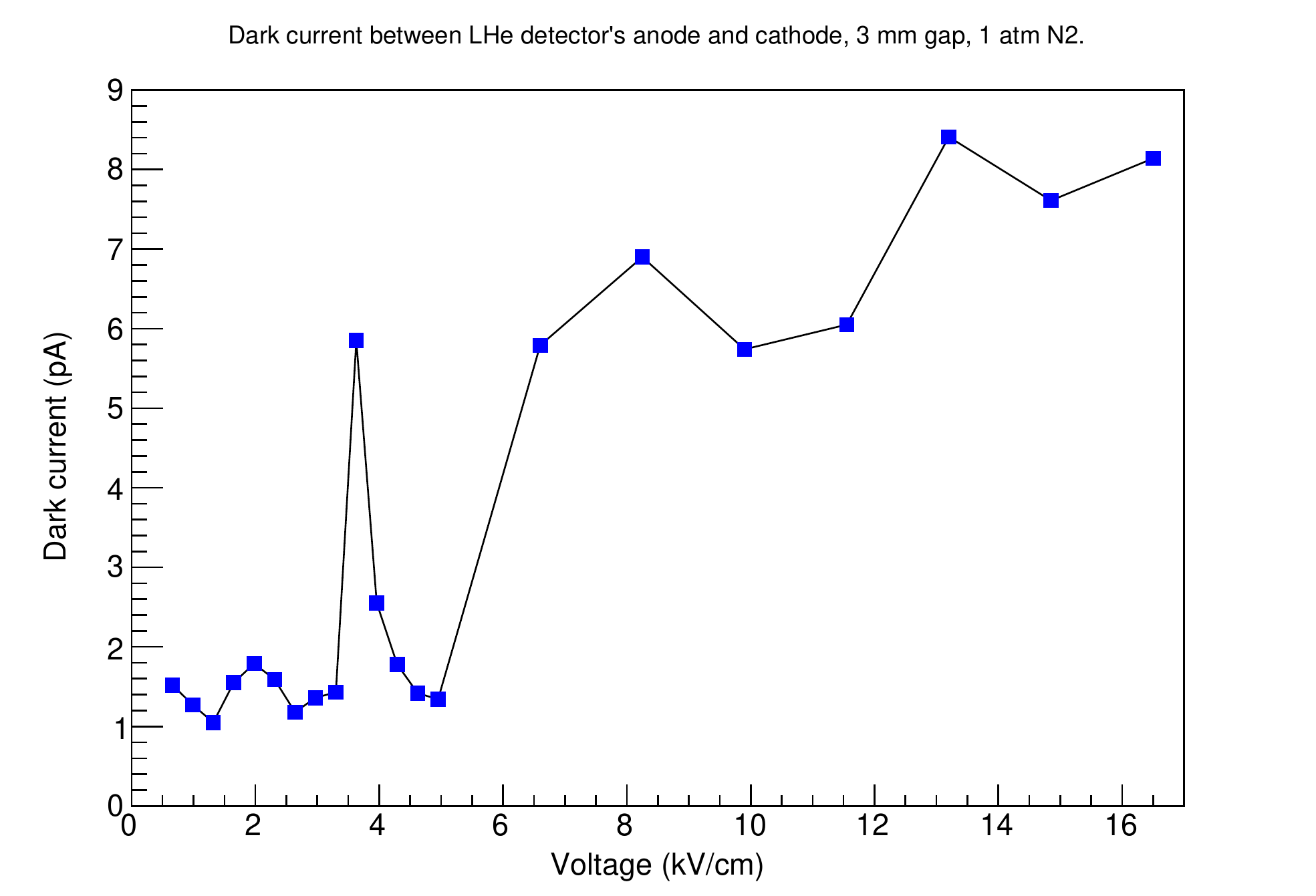}
		\caption{Fig.~\ref{fig_30gLHeDarkCurrentResults.b}. The dark current of the 30 g LHe cell, measured when filled with 1 atm nitrogen gas.}\label{fig_30gLHeDarkCurrentResults.b}
	\end{subfigure}
	\quad
		\begin{subfigure}[t]{3.1in}
		\centering
		\includegraphics[scale=0.35]{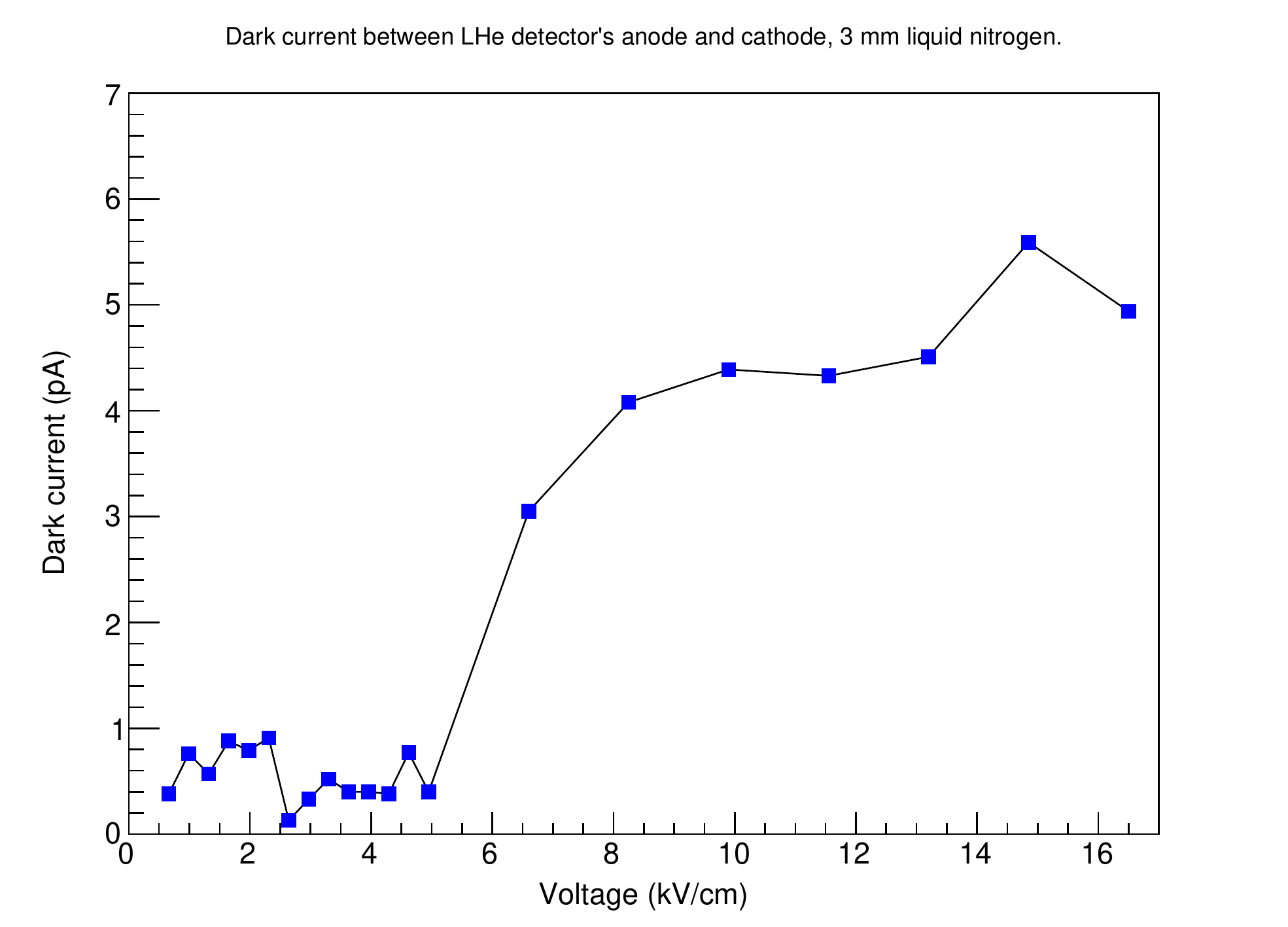}
		\caption{Fig.~\ref{fig_30gLHeDarkCurrentResults.c}. The dark current of the 30 g LHe cell, measured when filled with liquid nitrogen.}\label{fig_30gLHeDarkCurrentResults.c}	
		\end{subfigure}

	\caption{The dark current of the 30 g LHe detector is less than 10 pA when filled with vacuum, 1 atm nitrogen gas, and liquid nitrogen with a HV field up to 17 kV/cm.}\label{fig_30gLHeDarkCurrentResults}
\FloatBarrier
\end{figure}

Initially, we didn't implement a plasma gun for our cleaning protocol. Then, unexpectedly, we observed that dark current fluctuates up to 3 orders. A typical fluctuation phenomenon is that the initial current is tens of nA; after half an hour or more, it goes down to tens of pA. To mitigate such a fluctuation, we implemented a plasma gun. It indeed improved a lot but was still not stable enough. We later (mechanically) polished the surface of the electrodes. In the end, we measured $< 10 $ pA dark current with an HV field up to 17 kV/cm when the detector was filled with vacuum, 1 atm nitrogen gas, and liquid nitrogen, are shown in Fig~\ref{fig_30gLHeDarkCurrentResults}. 

We did not measure the dark current when filled with LHe yet because the liquefying plant has been out of work since September 2021. We will do that as long as the plant is applicable.

\captionsetup[subfigure]{labelformat=empty}

\section*{Summary for the ALETHEIA project}

TThe ALETHEIA project aims to hunt for low-mass DM. Filling with LHe, the program will implement the most convincing technology in the community of DM direct detection, TPC. Although lots of theoretical and experimental research have been launched to understand the particle characters of LHe, most of those experimental results focused on 100s keV$_{nr}$ energy or higher, which are out of the ROI of the proposed LHe TPC: $\sim$ 0.5 -10 keV$_{nr}$. Therefore, we have to launch a complete set of experimental tests to verify that an LHe TPC is suitable for low-mass DM hunting. 

The first 30 g LHe detector is designed and assembled at CIAE in Beijing, China. We successfully cooled the detector down to liquid helium temperature. The detector's dark current is less than ten pA for an external HV up to 17 kV/cm when filled with vacuum, 1 atm nitrogen gas, and liquid nitrogen.

\section*{Acknowledgement}
We thank the professors who flew to Beijing in Oct 2019 to participate in the DM workshop and reviewed the project: Prof. Rick Gaitskell, Prof. Dan Hooper, Prof. Jia Liu, Prof. Dan McKinsey, Dr. Takeyaso Ito, and Prof. George Seidel. We thank Prof. Weiping Liu for helping Junhui Liao settle down at CIAE. Junhui Liao would also thank the support of the ``Yuanzhang'' funding of CIAE to launch the 30 g LHe cell program.

\bibliography{ALETHEIA-Overleaf}

\end{document}